\documentclass[journal]{IEEEtran}
\ifCLASSINFOpdf
\else
\fi
\usepackage{booktabs}
\usepackage{multirow}
\usepackage{graphicx}
\usepackage{subfigure}
\usepackage{amsmath}
\usepackage{amssymb}
\usepackage{cite}
\usepackage{threeparttable}
\usepackage{adjustbox}

\usepackage{array}
\usepackage{xcolor}
\colorlet{blue}{black}
\colorlet{red}{black}

\begin{document}

\title{Conquering Data Variations in Resolution: A Slice-Aware Multi-Branch Decoder Network}


\author{Shuxin~Wang, %~\IEEEmembership{Member,~IEEE,}
Shilei~Cao, %~\IEEEmembership{Fellow,~OSA,}
Zhizhong~Chai, %~\IEEEmembership{Fellow,~OSA,}
Dong~Wei,
Kai~Ma,
Liansheng~Wang, ~\IEEEmembership{Member,~IEEE,}
and~Yefeng~Zheng,~\IEEEmembership{Senior~Member,~IEEE}% <-this % stops a space
\thanks{This work was supported by National Natural Science Foundation of China (Grant No. 61671399), the Fundamental Research Funds for the Central Universities (Grant No. 20720190012), the Key Area Research and Development Program of Guangdong Province, China (Grant No. 2018B010111001) and Science and Technology Program of Shenzhen, China (No. ZDSYS201802021814180). (Corresponding authors: Liansheng~Wang; Yefeng~Zheng.)}
\thanks{Shuxin~Wang, Zhizhong~Chai and Liansheng~Wang are with Department of Computer Science, Fujian Key Laboratory of Sensing and Computing for Smart City, Xiamen University,
Siming South Road, Xiamen 361005, China. Shuxin~Wang and Zhizhong~Chai contributed to this work when they were interns at Tencent (e-mail: shuxin\_icey@163.com, chai\_zhizhong1995@163.com, lswang@xmu.edu.cn).} % <-this % stops a space
\thanks{Shilei~Cao, Dong~Wei, Kai~Ma, and~Yefeng~Zheng are with Tencent Jarvis Lab, Malata Building, Kejizhongyi Road, Nanshan District, Shenzhen 518075, China (e-mail: eliasslcao@tencent.com, kylekma@tencent.com, donwei@tencent.com, yefengzheng@tencent.com).}
\thanks{Shuxin~Wang and Shilei~Cao contribute equally.}
}% <-this % stops a space


\markboth{Journal of \LaTeX\ Class Files}%
{Shell \MakeLowercase{\textit{et al.}}: Bare Demo of IEEEtran.cls for IEEE Journals}



\maketitle

\begin{abstract}
Fully convolutional neural networks have made promising progress in joint liver and liver tumor segmentation.
Instead of following the debates over 2D versus 3D networks (for example, pursuing the balance between large-scale 2D pretraining and 3D context), in this paper, we novelly identify the wide variation in the ratio between intra- and inter-slice resolutions as a crucial obstacle to the performance.
To tackle the mismatch between the intra- and inter-slice information, we propose a slice-aware 2.5D network that emphasizes extracting discriminative features utilizing not only in-plane semantics but also out-of-plane coherence for each separate slice.
Specifically, we present a slice-wise multi-input multi-output architecture to instantiate such a design paradigm, which contains a Multi-Branch Decoder (MD) with a Slice-centric Attention Block (SAB) for learning slice-specific features and a Densely Connected Dice (DCD) loss to regularize the inter-slice predictions to be coherent and continuous.
Based on the aforementioned innovations, we achieve state-of-the-art results on the MICCAI 2017 Liver Tumor Segmentation (LiTS) dataset.
Besides, we also test our model on the ISBI 2019 Segmentation of THoracic Organs at Risk (SegTHOR) dataset, and the result proves the robustness and generalizability of the proposed method in other segmentation tasks.
\end{abstract}


% \fi
\IEEEpeerreviewmaketitle

\section{Introduction}

\begin{figure}[!t]
\begin{center}
	\includegraphics[width=1.\columnwidth]{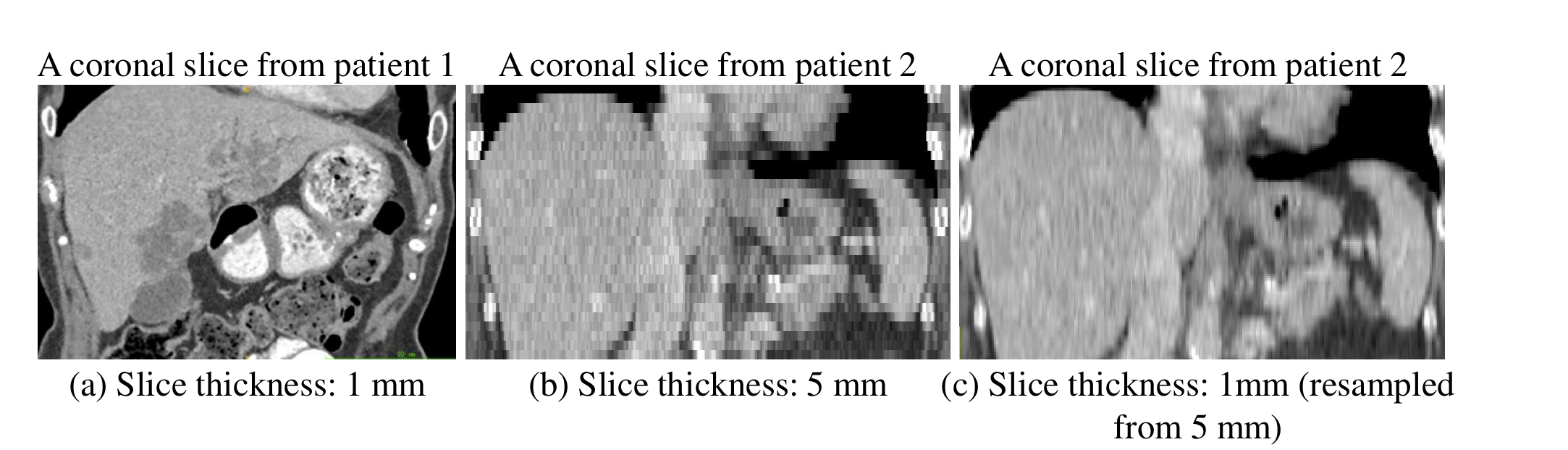}
\end{center}
\caption{An illustration of the wide variation in out-of-plane slice thickness with a fixed in-plane pixel spacing. A CT liver image with slice thickness of 1 $mm$ (a); another image with slice thickness of 5 $mm$ (b) and the corresponding resampled image of 1 $mm$ (c). We can observe that, comparing (a) and (b), CT images show apparent visual difference with different slice thicknesses; and comparing (a) and (c), CT images still show apparent visual difference (\textit{e.g.}, blurring) even after being resampled to the same slice thickness.}
\label{data_spacing}
\end{figure}

\IEEEPARstart{T}{he} liver is a vital organ in the human body as it is essential  for bile secretion and detoxifying harmful substances into urea. According to the global cancer statistics reported in 2018~\cite{bray2018global}, liver cancer is the sixth most frequently diagnosed cancer and the fourth leading cause of cancer death worldwide. The liver is also a common site for other metastatic cancer because of the rich blood supply \cite{meng2013huge}. In the current clinical routine, CT is the most frequently used imaging modality for radiologists and oncologists to make accurate hepatocellular carcinoma evaluation and treatment planning \cite{gaiani2004usefulness}. Nevertheless, outlining the liver and liver tumor in CT slice-by-slice is a time-consuming task and prone to annotator variations. Therefore, a standardized and automatic segmentation method is highly desirable to enable efficient delineation of liver and liver tumor contours in practice.

While remarkable performance on liver and liver tumor segmentation was recently reported with the development of deep learning~\cite{chlebus2017neural,vorontsov2018liver,li2018h, liu2018automatic,liu20183d,deng2019dynamic,han2017automatic}, a few challenges deserve wide attention of the community. Firstly, to achieve a good generalization of the established model, data are usually collected from various clinical sites. However, due to the variations in equipment manufacturers, physical parameters, scanning protocols, and reconstruction methods, the voxel resolution of CT images from multiple centers suffers from wide variations, especially along the out-of-plane direction. Taking the Liver Tumor Segmentation (LiTS) dataset \cite{bilic2019liver} as an example, its in-plane pixel spacing ranges from $0.56$ $mm$ to $1$ $mm$, whereas its out-of-plane slice thickness ranges from $0.45$ $mm$ to $6$ $mm$. Secondly, liver and liver tumor segmentation in CT images is a task to assign coherent semantic masks to the full volume, rather than to individual slices, which implies that objects in adjacent slices of a volume usually have intrinsic correlations in terms of context, shape, and location.
In conclusion, handling the information asymmetry as a result of the inconsistent in- and out-of-plane resolutions and ensuring the inter-slice segmentation consistency are two critical issues to the liver and liver tumor segmentation.

There were some remarkable deep learning works for liver and liver tumor segmentation, which can be roughly classified as 2D \cite{chlebus2017neural, vorontsov2018liver}, 2.5D \cite{han2017automatic} and 3D \cite{liu2018automatic,deng2019dynamic} methods. Standard 2D methods can learn abundant deep semantics by employing deeper neural networks with large quantities of training samples. However, losing inter-slice information makes it hard to learn a smooth segmentation map along the out-of-plane direction. On the contrary, 2.5D and 3D methods can exploit 3D context information and thus learn more meaningful feature representations to maintain 3D coherence. However, most of the existing methods underestimate the impact of the differences between the in-plane pixel spacing and out-of-plane slice thickness, and simply resample the input to a fixed in-plane pixel spacing and out-of-plane slice thickness.
Although the resampling process can relieve the resolution anisotropy problem, it does not bring extra information to the dimension(s) being finely \emph{interpolated}, while some additional artifacts may be introduced (see Fig.~\ref{data_spacing}).
Besides, 3D methods usually suffer from high computational cost and GPU memory consumption, which may hinder application in practice.

In this paper, we propose a novel 2.5D Slice-Aware Multi-Branch Decoder (SAMBD) network which utilizes not only the large-scale 2D pre-training but also 3D contextual information based on the observation of variations in data.
SAMBD is focused on learning discriminative slice-specific features, and instantiates a slice-wise multi-input multi-output architecture for this goal.
The core components of SAMBD can be summarized in three parts:

(i) \textbf{A mutual encoder.} We employ the Xception \cite{chollet2017xception} model with its weights pre-trained on natural images (except for the first layer) to simultaneously extract deep semantics for multiple input slices with discriminative feature initialization, which captures the local volumetric information at different semantic levels.
Motivated by DeepLabv3+ \cite{chen2018encoder}, an Atrous Spatial Pyramid Pooling (ASPP)~\cite{chen2018encoder} module is adopted to fuse features of different scales.

(ii) \textbf{A Multi-branch Decoder (MD) with Slice-centric Attention Block (SAB).}
Since the multiple input slices are indiscriminately processed by the encoder and the slice-specific information is thus scrambled, we design an MD to explicitly re-establish discriminative features for each slice by fully exploiting the intra- and inter-slice information learned by the encoder.
To further strengthen the discriminative power of each slice, we propose and embed an SAB into the MD, which is implemented with the widely adopted attention mechanism \cite{oktay2018attention,shao2019attentive}.
All these designs are centered around learning the best features for each slice, thus avoiding directly processing asymmetric intra- and inter-slice information.

(iii) \textbf{A Densely Connected Dice (DCD) loss.}
Based on the assumption that target objects lying in successive slices should have consistent labels, we propose a DCD loss to regularize the inter-slice predictions to be more coherent in the label space, where the intra- and inter-slice constraints can be jointly optimized.

% In order to intuitively prove the effectiveness of such slice-aware design, we also propose a 3D variant of the 2.5D multi-branch decoder. Compared to the 3D slice-insensible decoder, the 3D slice-aware decoder improves the segmentation accuracy.

% Different from previous methods, the proposed network preserves their strong generalization capability for spatial information while naturally exploiting the inter-slice information for more effective modeling of anisotropic images.

% Based on the aforementioned innovations, we train a deep CNN in an end-to-end fashion and achieve the state-of-the-art result on the MICCAI 2017 Liver Tumor Segmentation~(LiTS) dataset \cite{bilic2019liver}. Moreover, we conduct thorough ablation studies to examine the effectiveness of each component.
% Additionally, to prove the robustness and generalizability of the proposed method, we also test our model on the ISBI 2019 Segmentation of THoracic Organs at Risk (SegTHOR) dataset \cite{trullo2019multiorgan}, and the result is very competitive compared with other methods.

In summary, the contributions of this work can be summarized as four-fold:
\begin{itemize}
	\item Instead of considering the debates over 2D versus 3D networks, in this paper, we identify the wide variation in the ratio between the intra- and inter-slice resolutions as an important obstacle to the performance.
	\item Observing the variations in data, we propose a 2.5D encoder-decoder network with a multi-input and multi-output structure, featuring a novel slice-aware multi-branch decoder with a slice-centric attention block which not only utilizes the large-scale 2D pre-training but also 3D contextual information for learning discriminative features for each separate slice.

	\item An auxiliary loss function is proposed to strengthen the inter-slice correlations and regularize the inter-slice predictions to be more coherent.

	\item We mainly evaluate our method on the CT volumes for liver and liver tumor segmentation provided by LiTS and the result outperforms other methods. Besides, extended validation is conducted on the ISBI 2019 Segmentation of THoracic Organs at Risk (SegTHOR) dataset and the result is competitive.
\end{itemize}

The remainders of this paper are organized as follows. We review the related work in Section~\ref{related_work} and elaborate on the proposed method in Section~\ref{method}. We present experiments and results in Section~\ref{experiments}, a discussion in Section~\ref{discussion}, and finally draw the conclusions in Section~\ref{conclusion}.

\begin{figure*}[!t]
\begin{center}
	\includegraphics[width=0.7 \linewidth]{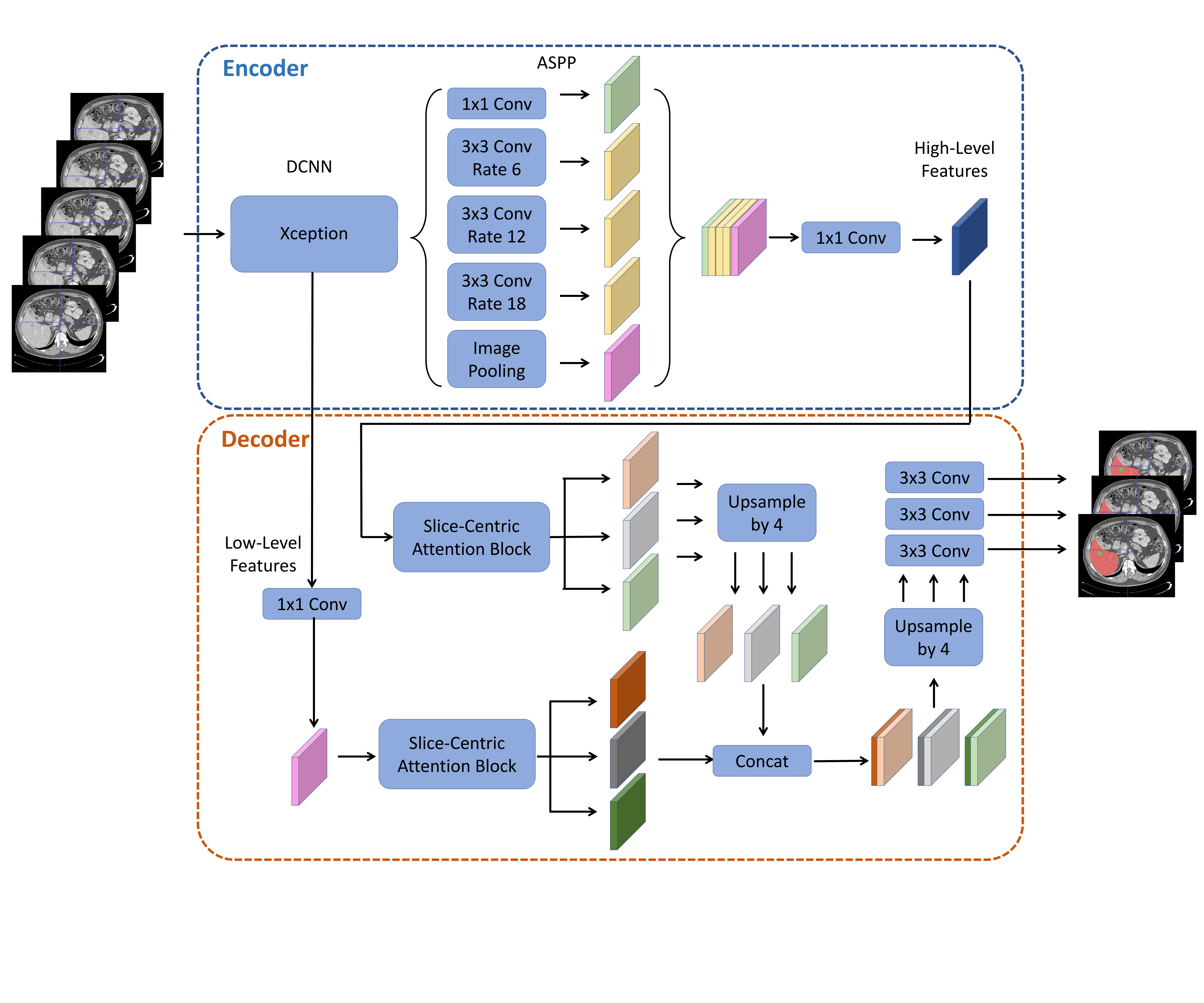}
\end{center}
\caption{Schematic view of our proposed framework with a set of five input slices ($C_\mathrm{in}$=5) and the corresponding three central slice predictions ($C_\mathrm{out}$=3). A multi-branch decoder with a slice-centric attention block is proposed to gradually and explicitly re-establish discriminative features for each slice by fully exploiting intra- and inter-slice information learned by the encoder.}
\label{framework}
\end{figure*}

\section{Related Works} \label{related_work}
\subsection{Debates over 2D versus 3D Networks}
Prior studies chose either 2D networks for the benefits of 2D pretraining and large-scale slice-wise training sets or alternatively 3D networks for native 3D representation learning \cite{yang2019reinventing,yu2019thickened}.

Recently, the LiTS challenge was organized to benchmark the performance of different automatic algorithms for liver and liver tumor segmentation, where the top-scoring methods were dominated by Fully Convolutional Networks (FCNs) \cite{long2015fully}.
(i) \textbf{2D/2.5D networks}. Vorontsov \textit{et al.} \cite{vorontsov2018liver} segmented liver and liver tumor with two FCNs, which were connected in tandem and trained together end-to-end, using a 2D axial slice as input.
To capitalize on the complementary information between a few adjacent slices, Han \cite{han2017automatic} proposed a 2.5D model, which combined the long-range connection of U-Net~\cite{ronneberger2015u} and the short-range connection of ResNet~\cite{he2016deep}.
Other noteworthy works~\cite{wang2018triplanar,chung2019liver,chlebus2017neural} attempted to use triplanar networks to learn generalized features from the axial, coronal, and sagittal planes.
(ii) \textbf{3D networks}.
There were some works employing 3D convolution to mine 3D context information. For example, Liu \textit{et al.} \cite{liu2018automatic} implemented an improved 3D U-Net equipped with dilated convolutions and separable convolutions to segment livers. Deng \textit{et al.} \cite{deng2019dynamic} proposed dynamic regulation of level-set parameters using 3D CNN for liver tumor segmentation.
(iii) \textbf{Hybrid approaches}. To simultaneously take advantage of the merits of 2D and 3D networks, Li \textit{et al.} \cite{li2018h} proposed a novel hybrid densely connected U-Net named H-DenseUNet, which consists of a 2D DenseUNet for efficiently extracting intra-slice features and a 3D counterpart for hierarchically aggregating volumetric contexts for better liver and tumor segmentation.
With similar motivation, Zhang \textit{et al.} \cite{zhang2019light} proposed a light-weight hybrid convolutional network to segment the liver and liver tumors with an encoder-decoder structure, in which 2D convolutions used at the bottom of the encoder decreases the complexity and 3D convolutions used in other layers explore both in- and out-of-plane information.

Despite the remarkable performance achieved by the aforementioned methods, they underestimated the impact of the inconsistency between the pixel spacing and slice thickness of 3D volumetric data.
Our work begins with the basic observation of the data variations which motivates us to extract discriminative features for each slice, based on the fact that the intra-slice information is more coherent due to the uniform in-plane pixel spacing than inter-slice information.

\subsection{Approaches to Conquering Data Variations in Resolution}
Intuitively, resampling to a unified resolution may be a solution to the data variation problem;
however, the resampled images suffer from different information densities along the dimensions, and the resampling operation cannot guarantee the validity of the interpolated information as also mentioned in \cite{li2018h}.

It is known that thin slice thickness (less than 2.5 $mm$) results in better performance both for the human reader and computer-aided diagnosis (CAD) systems; however, CT scans with thicker slice thickness (greater than 2.5 mm) are widely used in clinical setting mainly because of the efficiency in terms of reading time and storage \cite{baeresidual}.
Some works \cite{baeresidual,ge2019stereo} explored to reduce the slice thickness of CT scans from thick to thin.
For example, Bae \textit{et al.} \cite{baeresidual} proposed a 2.5D image super-resolution (SR) network based on fully residual convolutional neural networks (CNN) for dense slice reconstruction.
Ge \textit{et al.} \cite{ge2019stereo} proposed a residual voxel-wise generative adversarial network, which densely reconstructed slices into a thin thickness (1 $mm$) and meanwhile denoised the CT images into the more readable pattern, from the widely accessible low-dose thick CT.
Although we can adopt such techniques to reconstruct CT images into thin slices of a unified resolution, extra computational costs would be inevitably incurred.

\section{Methodology}\label{method}
In this section, we present the details of the proposed Slice-Aware Multi-Branch Decoder (SAMBD) network. The network architecture is depicted in Fig. \ref{framework}. In general, the network design adopts the standard encoder-decoder structure, where the encoder takes a stack of adjacent slices as input and outputs compact feature representations. Meanwhile, the decoder restores the feature maps to the original resolution by fusing features from different levels of the encoder and outputs the label predictions for the central slices. Here, we use $C_\mathrm{in}$ and $C_\mathrm{out}$ to denote the numbers of input slices and output slice predictions ($C_\mathrm{out}=C_\mathrm{in}-2$ in this work), respectively. No segmentation output is generated for the top or bottom slice since there is not enough context for these boundary slices.

\subsection{Encoder} \label{A}
Inspired by the success of the design of DeepLabv3+ \cite{chen2018encoder}, the encoder consists of a modified Xception~\cite{chollet2017xception} structure as the backbone and an ASPP~\cite{chen2017rethinking} module.
By adopting depthwise separable convolution, the Xception model \cite{chollet2017xception} achieves improvement in terms of both speed and accuracy in semantic segmentation. We employ it as our network backbone for its strong feature representation power and small model size. Since the original Xception model processes a three-channel color image, we modify the input channel number to $C_\mathrm{in}$ to jointly process adjacent slices. The ASPP module can potentially improve the segmentation performance by involving different sampling rates and enlarging effective field-of-views, thus capturing target objects as well as context information at different scales. We adopt it here to cover the tumors of various sizes.
We initialize the encoder with the weights pre-trained on PASCAL VOC 2012 \cite{pascal-voc-2012} provided by the official implementation of DeepLabv3+ \cite{chen2018encoder}.

\subsection{Multi-branch Decoder}\label{B}
\begin{figure}[!t]
\color{blue}
\begin{center}
	\includegraphics[width=0.9 \columnwidth]{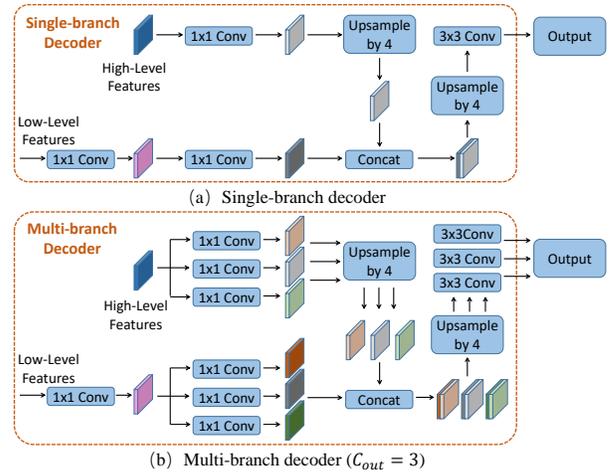}
\end{center}
\caption{\color{red}Single-branch decoder vs. our proposed multi-branch decoder. The single-branch decoder only generates one-slice output while the multi-branch decoder simultaneously generates $C_\mathrm{out}$ (equals three in this figure) outputs. Note that the low-level features are from the first residual block of Xception \cite{chollet2017xception}, while the high-level features are the output of the ASPP module \cite{chen2018encoder}.}
\label{fig:single-branch-vs-multi-branch}
\end{figure}

In the decoder design, a multi-branch structure is proposed to distill the slice-specific information from the encoded volumetric features.
Formally speaking, the decoder structure used in 2D or 2.5D FCNs from previous studies can be seen as a single-branch decoder (as illustrated in Fig. \ref{fig:single-branch-vs-multi-branch}(a)), where only one central slice is predicted. This single-branch decoder balances the inherent tension between semantics and location, enables precise localization, and produces semantically meaningful predictions from the rich context. However, there are three problems that deserve attention. Firstly, anisotropic volumes have inconsistent information densities along different dimensions. When the slice thickness is much larger than pixel spacing, the single-branch decoder would learn mismatched information along different axes. Secondly, the encoder extracts features by simply fusing intra- and inter-slice information together using isotropic operators, and it fails to extract slice-specific features and loses the inter-slice structural information. Thirdly, the segmentation maps predicted by standard 2D or 2.5D approaches suffer from semantic inconsistency in neighboring slices, since each segmentation map of a slice is separately predicted by one forward inference.

To address the above problems, we design the decoder to have the same number of branches as the number of output slice predictions, and each branch shares the same structure with the single-branch decoder introduced in~\cite{chen2018encoder}. The design of our multi-branch decoder is motivated by the fact that the slice-specific information is scrambled through the encoder and we should re-establish them in the decoder with the rich volumetric information provided by the encoder. In this sense, we can explicitly associate one particular branch with one slice, thus bringing more room for improvement by exploiting structure prior between slices.

As shown in Fig. \ref{fig:single-branch-vs-multi-branch}(b), we take the low-level features (from the first residual block of Xception) and high-level features (from the outputs of the ASPP module) as the input of the multi-branch decoder (a design similar to DeepLabv3+ \cite{chen2018encoder}). For low-level features, we first employ a $1\times1$ convolution on them to reduce the number of channels, since too many channels in low-level features would outweigh the importance of the rich encoder features and make the training harder~\cite{chen2018encoder}. Then, $C_\mathrm{out}$ number of $1\times1$ convolutions are conducted on the outputs to explicitly associate one particular branch with one slice. For high-level features, we similarly adopt $C_\mathrm{out}$ number of $1\times1$ convolutions on them and upsample their outputs by a bilinear upsampling layer (with a factor of four) to make the scale consistent with the low-level branch. After that, we get the refined slice-specific features that cover information from both low- and high-level semantics of the encoder. A concatenation operation is then slice-wisely conducted on them to merge the multi-scale semantics. Finally, we adopt another $C_\mathrm{out}$ number of bilinear upsampling operations with a rate of four and $3\times3$ convolutions to form the final segmentation outputs.

\subsection{Slice-Centric Attention Block (SAB)}\label{C}
Instead of directly splitting the encoded feature maps into slice-specific parts as shown in Fig. \ref{fig:single-branch-vs-multi-branch} (b), we propose a novel SAB to strengthen the discrimination of slice-specific features by considering the inter-slice correlations with an attention mechanism.
The motivation of this block is based on the observation that the $1\times1$ convolutions employed in the multi-branch decoder to extract slice-specific information are overly simplistic, which is hard to effectively extract discriminative information for individual slices.
In contrast, the attention mechanism steers the allocation of slice-specific semantic features towards the most informative components for each output slice and explores the inter-slice correlations, hence improving the performance in learning slice-specific features.
For implementation, we embed the proposed SAB into the multi-branch decoder in both the low- and high-level decoding paths as shown in Fig. \ref{framework}.
Note that the two SABs do not share weights since they face different contexts in two different scales.

\begin{figure}[!t]
\centering
\includegraphics[width=1 \columnwidth]{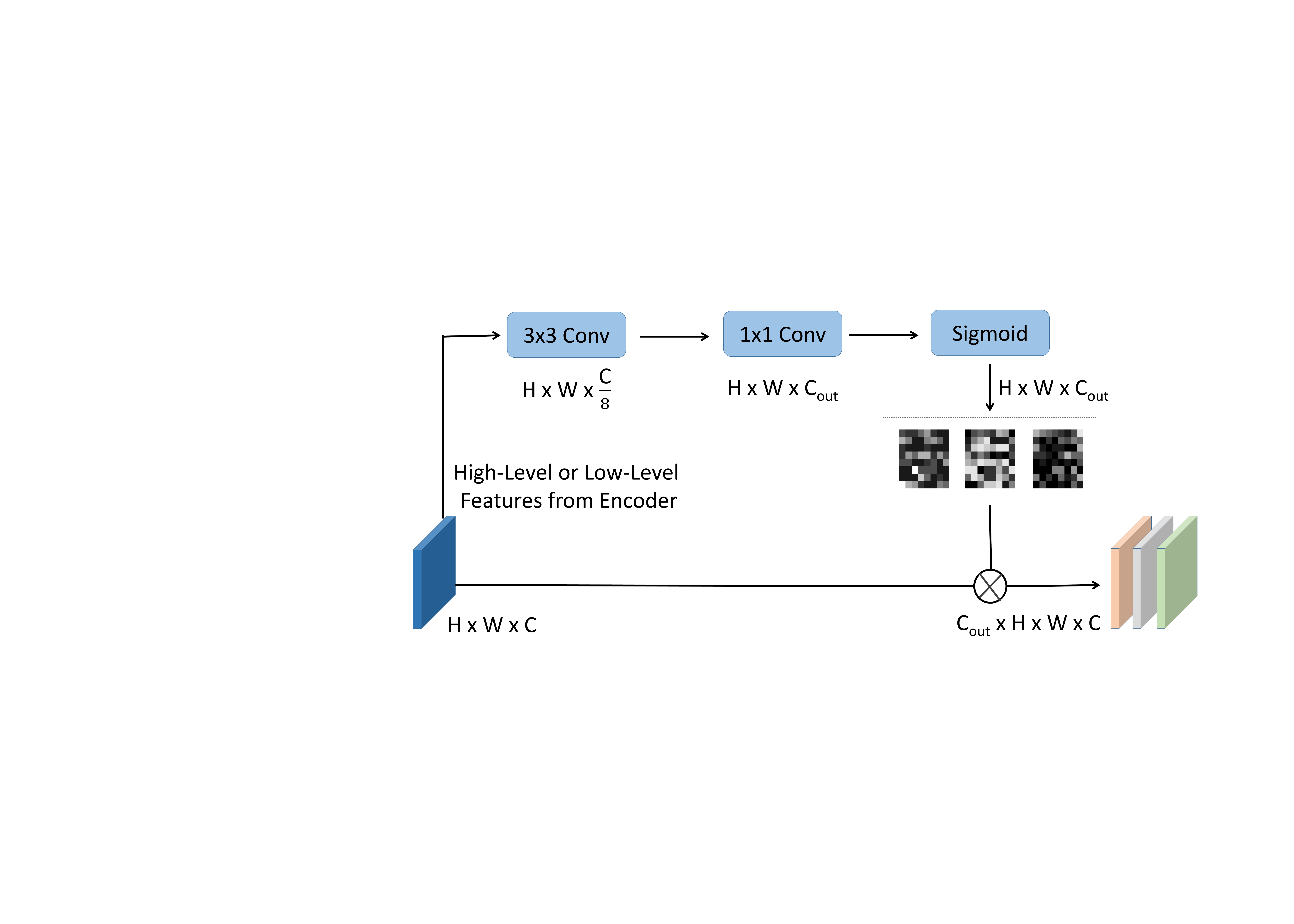}
\caption{Illustration of the proposed slice-centric attention block ($C_\mathrm{out} = 3$).
It strengthens the discriminative power of the multi-branch decoder in learning slice-specific features with an attention mechanism. C is the channel number of the high-level or low-level features, which is the same as the original DeepLabv3+.}
\label{fig:slice_disentanglement_block}
\end{figure}

The attention mechanism is widely adopted in medical image applications, such as pancreas segmentation \cite{oktay2018attention} and universal lesion detection \cite{shao2019attentive}.
Our work innovatively explores its usage on the problem of extracting discriminative features for each slice.
Fig. \ref{fig:slice_disentanglement_block} shows the technical implementation of the proposed SAB.
The volumetric features first pass through a $3 \times 3$ convolution layer, with one-eighth of channel numbers of the input feature map.
Then, $C_\mathrm{out}$ $1 \times 1$ convolution layers and sigmoid functions are applied to generate $C_\mathrm{out}$ weight maps of size $H\times W \times C_\mathrm{out}$.
These weight maps can be seen as an attention mechanism for each branch attending to the key slice-specific information from the abundant features. The learned weight maps are then separately multiplied by the input features to extract slice-specific features.
As we will demonstrate in the experiments, although conceptually simple, the proposed SAB is effective in strengthening the discriminative power for each slice in feature learning.

\subsection{Loss Function}\label{E}
The Dice loss introduced in~\cite{milletari2016v} is commonly used to address the class imbalance problem between foreground and background classes. As defined below in Eq. (\ref{eq:dice1}), the Dice loss with three classes (\textit{i.e.,} background, liver, and liver tumor) is formulated as:
\begin{equation}
\label{eq:dice1}
L_{Dice}=-\sum_{c=1}^{3}\sum_{m=1}^{C_\mathrm{out}}{\frac{2\sum_{i=1}^{V}p_{m,i}^{c}g_{m,i}^{c}}{\sum_{i=1}^{V}(p_{m,i}^{c})^{2}+\sum_{i=1}^{V}(g_{m,i}^{c})^{2}}},
\end{equation}
where $p_{m,i}^{c}$ denotes the predicted probability of voxel $i$ in the $m^{th}$ slice belonging to class $c$; $g_{m,i}^{c}$ denotes the corresponding ground truth; and $V$ is the number of voxels in each slice.

As aforementioned, slice-aware design brings room for improvement by exploiting structure prior between slices. We thus propose a regularization term as an additional loss to improve the coherence between neighboring slices in the label space. Specifically, we use the union of the two slices in the prediction results and the union of the corresponding two slices in the ground truth to calculate the pairwise Dice loss, which is denoted as
\begin{equation}
\label{eq:dice2}
\begin{split}
P_{m,n}=-\sum_{c=1}^{3}{\frac{2\sum_{i=1}^{V}(p_{m,i}^{c}+p_{n,i}^{c})(g_{m,i}^{c}+g_{n,i}^{c})}{\sum_{i=1}^{V}(p_{m,i}^{c}+p_{n,i}^{c})^{2}+\sum_{i=1}^{V}(g_{m,i}^{c}+g_{n,i}^{c})^{2}}},
\end{split}
\end{equation}
where $P_{m,n}$ denotes the pairwise Dice loss of the $m^{th}$ slice and $n^{th}$ slice along out-of-plane direction; $p_{m,i}^{c}$ and $p_{n,i}^{c}$ denote the predicted probability of voxel $i$ belonging to class $c$ in the $m^{th}$ slice and $n^{th}$ slice, respectively;  $g_{m,i}^{c}$ and $g_{n,i}^{c}$ denote the corresponding ground truth. To further supplement inter-slice information flow and improve inter-slice coherence, we calculate the pairwise Dice loss in a dense way, where each slice is coupled with multiple nearby slices to calculate multiple Dice losses. We name the new loss as the Densely Connected Dice (DCD) loss. \textcolor{blue}{Since the interaction of two slices decreases with increasing distance, for each paired slices $m$ and $n$ ($n > m$), we add a weight $w_{m,n}=1/(n-m)$.} The DCD loss is defined as:
% \begin{equation}
% \label{eq:dice3}
% \begin{split}
% L_{DCD}&=\sum_{m=1}^{C_\mathrm{out}-1}\sum_{n=m+1}^{C_\mathrm{out}}P_{m,n}w_{m,n} \\
%  &=\sum_{m=1}^{C_\mathrm{out}-1}\sum_{n=m+1}^{C_\mathrm{out}}P_{m,n}\times{\frac{1}{n-m}}
% \end{split}
% \end{equation}
\begin{equation}
\label{eq:dice3}
L_{DCD}=\sum_{m=1}^{C_\mathrm{out}-1}\sum_{n=m+1}^{C_\mathrm{out}}w_{m,n}P_{m,n}.
\end{equation}

The final loss function is composed of a weighted combination of the Dice loss and the proposed DCD loss:
% \begin{equation}
% \label{eq:dice4}
% \begin{split}
% L&=L_{Dice}+ \lambda\times L_{DCD} \\
% &=L_{Dice}+\frac{C_\mathrm{out}}{\sum_{m=1}^{C_\mathrm{out}-1}\sum_{n=m+1}^{C_\mathrm{out}}w_{m,n}}\times L_{DCD}.
% \end{split}
% \end{equation}
\begin{equation}
\label{eq:dice4}
L=L_{Dice}+ \lambda\times L_{DCD},
\end{equation}
where \textcolor{blue}{we define $\lambda=C_\mathrm{out} / (\sum_{m=1}^{C_\mathrm{out}-1}\sum_{n=m+1}^{C_\mathrm{out}}w_{m,n})$ to balance the importance of the intra-slice semantic constraint and inter-slice smoothness.
$\lambda$ is such designed that its denominator normalizes the sum of the weights $w_{m,n}$ in Eq. (\ref{eq:dice3}) to one, whereas its numerator strengthens the supervision with more output slices.
Empirical parameter tuning is likely to yield better results;
in this paper, however, we would like to present a generally useful regularizer that can be safely applied to other segmentation tasks without any parameter tuning.
We find that the presented design of $\lambda$ performs well as validated by the superior performance on two publicly available datasets in the experiments.}

\begin{table}[!t]
	\color{blue}
	\caption {Comparison of our proposed SAMBD ($C_\mathrm{in}=7,\ C_\mathrm{out}=5$) with DeepLabv3+ \cite{chen2018encoder} and H-DenseUNet \cite{li2018h} in terms of parameters and FLOPs.} \label{tab_complexity}
	\centering

	\begin{tabular}{c|ccc}
   \toprule

	 &  DeepLabv3+ & SAMBD & H-DenseUNet \\
	\midrule
	Parameters (M) & 41.06 & 41.26 & 61.44  \\
	FLOPs (G) & 0.83 & 0.84 &\color{red} 2841.6  \\
	\bottomrule
	\end{tabular}

\end{table}

{\color{blue}
\subsection{Model Complexity} \label{F}
Due to the effective design of the multi-branch decoder and slice-centric attention block, the parameters and {\color{red}FLOPs (undefined)} of our proposed SAMBD with $C_\mathrm{in}=7,\ C_\mathrm{out}=5$ are very competitive with DeepLabv3+ \cite{chen2018encoder}, and markedly superior to H-DenseUNet \cite{li2018h}.
As shown in Table \ref{tab_complexity},  our method only brings 0.49\% extra parameters and 1.2\% extra FLOPs compared to DeepLabv3+, and only incurs 67.2\% and 0.03\% of the parameters and FLOPs of H-DenseUNet, respectively.}

\section{Experiments}\label{experiments}
In this section, we evaluate our approach on the LiTS \cite{bilic2019liver} dataset to demonstrate the robustness and generalization capability, compared to several state-of-the-art segmentation methods. Extended experiments have also been performed on the ISBI 2019 SegTHOR dataset \cite{trullo2019multiorgan} to validate the generalization capability to other human organs. We implement all the experiments with \textit{Keras}~\cite{chollet2015keras} using three NVIDIA GeForce GTX 1080 GPUs. Stochastic gradient descent with momentum (0.9) is used to update the weights of the network. The initial learning rate is set to 0.001 and multiplied by 0.9 after each epoch. We train the network for 80 epochs.
%The number of epochs is 80 in total.

\subsection{Experiments on LiTS}
\subsubsection{LiTS}
\begin{figure}[!t]
	\begin{center}
		\includegraphics[width=0.8 \columnwidth]{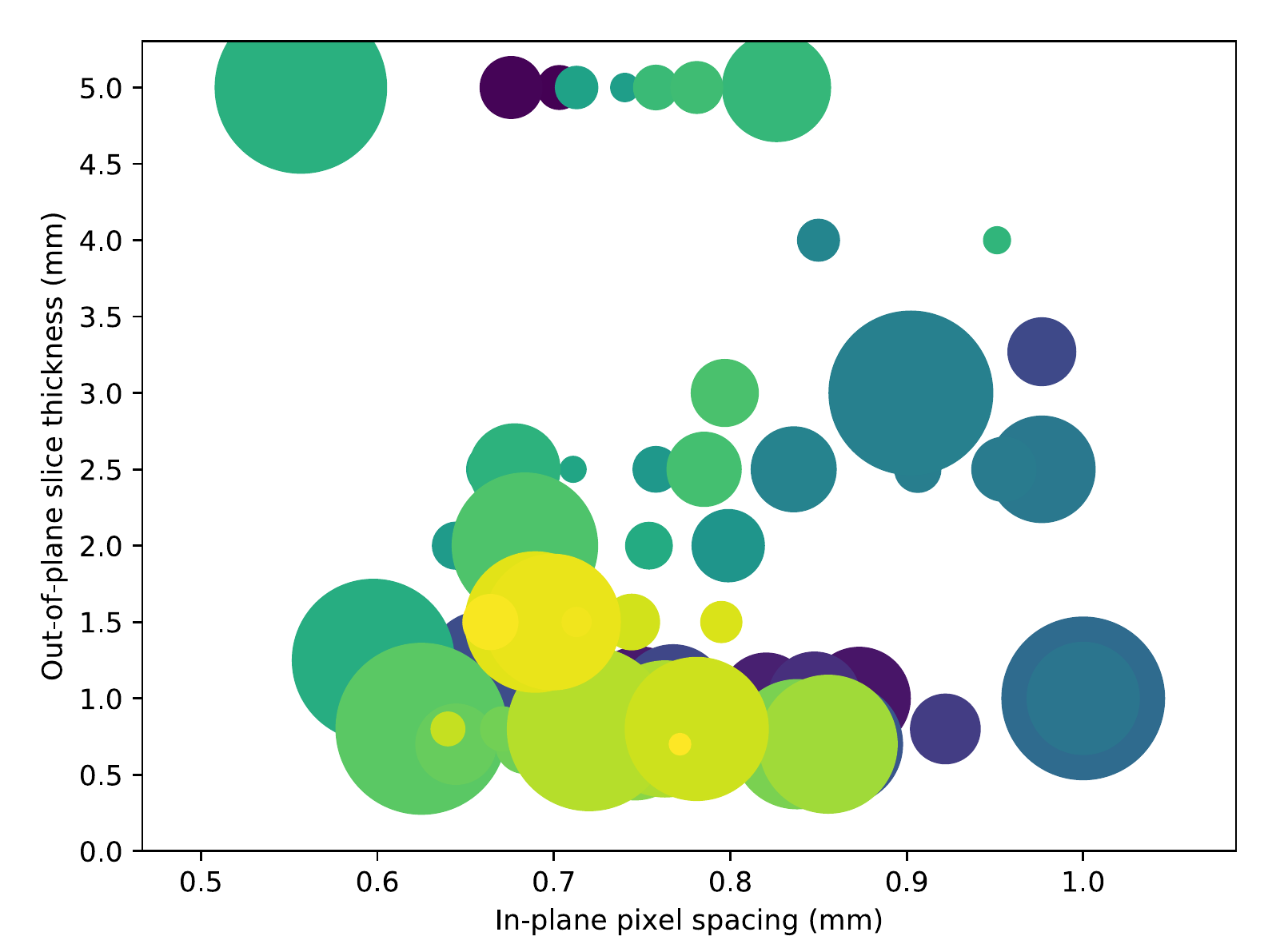}
	\end{center}
	\caption{The high variations of the LiTS training set. We can observe the large inconsistency between the in-plane pixel spacing (horizontal axis) and out-of-plane slice thickness (vertical axis). Besides, the variations in tumor size (shown in circles with different sizes) are also high.}
	\label{spacing}
\end{figure}

\begin{figure}[!t]
	\begin{center}
		\includegraphics[width=1.\columnwidth]{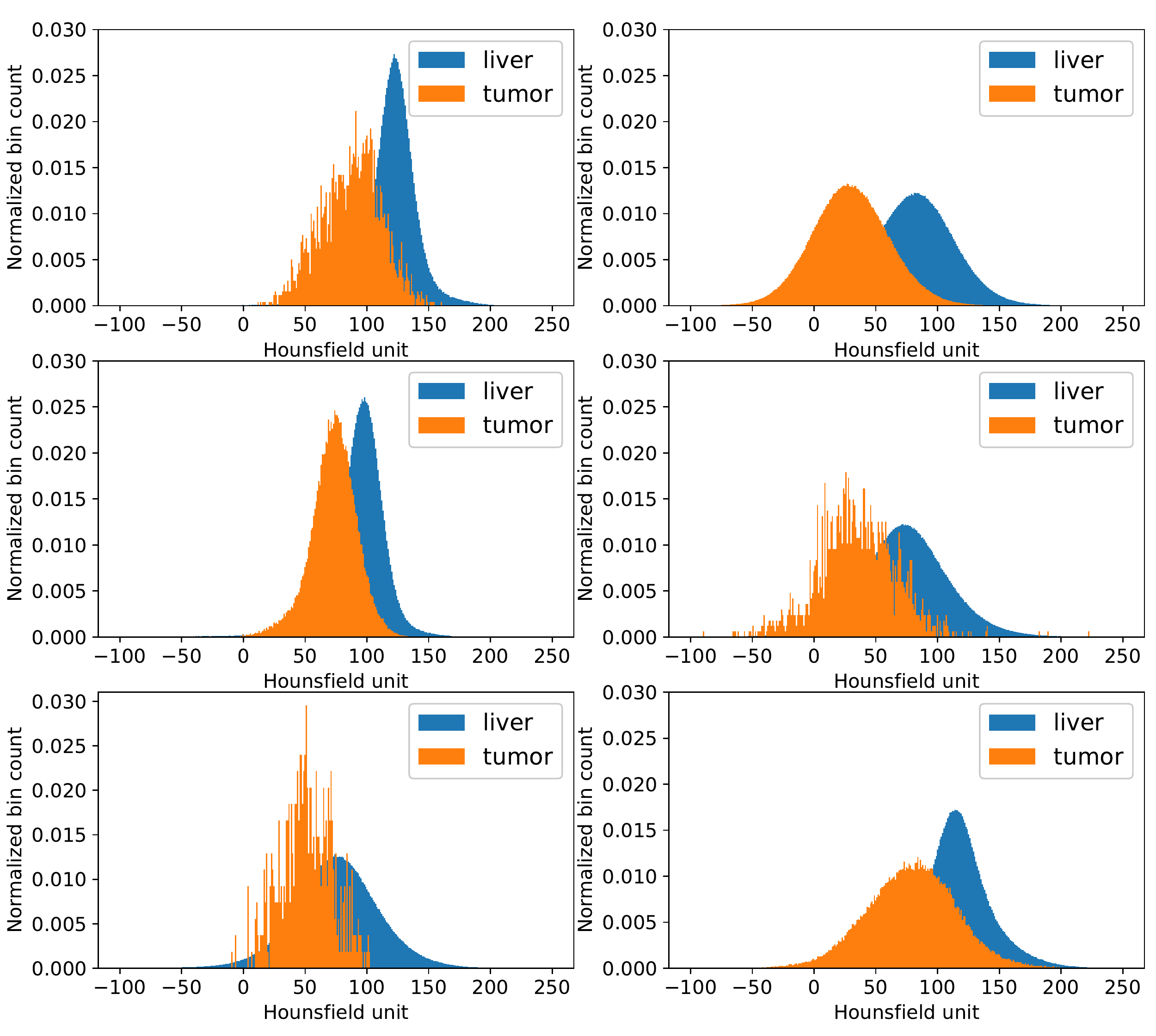}
	\end{center}
	\caption{Examples depicting the heterogeneity of CT scan contrast in liver and liver tumor areas. The horizontal axis represents the Hounsfield unit values of CT scans, and the vertical axis represents the proportion of voxels falling into intervals with different Hounsfield unit values.}
	\label{contract}
\end{figure}

\textcolor{blue}{The LiTS dataset}\footnote{\color{blue}https://competitions.codalab.org/competitions/17094} \textcolor{blue}{\cite{bilic2019liver}} is a publicly available liver tumor dataset consisting of 201 contrast-enhanced abdominal CT scans collected from various clinical sites over the world. The dataset was originally split into a training set (131 scans) and a test set (70 scans), where only the training set was publicly released with accurate liver and tumor masks. As aforementioned, the LiTS dataset suffers from apparent inconsistency between the in-plane pixel spacing and out-of-plane slice thickness (see Fig. \ref{spacing}).
% Its pixel spacing ranges from $0.56$ $mm$ to $1$ $mm$, while its out-of-plane slice thickness ranges from $0.45$ $mm$ to $6$ $mm$ (see Fig.\ref{spacing}).
% The large diversity of slice thickness leads to differences in conveying information along different direction as shown in Fig.\ref{data_spacing}.
Furthermore, high varieties and complexities exist for livers and liver tumors, including the location, size, and shape. Besides, the heterogeneity in liver and liver tumor contrast is very large between subjects, as shown in Fig. \ref{contract}.

\subsubsection{Implementation Details and Evaluation Metrics}
For image pre-processing, we unify the volume orientations and truncate the image intensity values of all scans outside the range of $[-200, 250]$ Hounsfield Unit (HU) to ignore irrelevant image details. Since the slice thickness varies greatly between subjects, we resample scans with slice thickness greater than $1$ $mm$ to $1$ $mm$ in both training and inference phases. We preserve the original slice thickness for patients whose slice thickness is less than $1$~$mm$ to leverage the original high-resolution spatial information. We do not unify the in-plane pixel spacing by the resample operation since the variation is relatively small and interpolation often introduces artifacts, which may offset the performance gain from resolution normalization. To alleviate the overfitting problem, we conduct data augmentation in the training phase; concretely, we first apply random scaling (from $0.8$ to $1.2$) to all training data, and then randomly crop a $256 \times 256 \times C_\mathrm{in}$ subregion as the input to the network. For post-processing during the inference phase, we take the largest connected component as the liver segmentation and remove liver tumor predictions outside the liver region.

\begin{table*}[!t]
	\caption {Liver and tumor segmentation results of an ablation study on the LiTS validation dataset.} \label{tab1}
	\centering
	\begin{adjustbox}{width=\textwidth}

	\begin{threeparttable}
		\begin{tabular}{c c|c|c c c|c|c|c|c|c|c}
			\toprule[2pt]
			\multirow{2}{*}{\bf{$C_\mathrm{in}$}\tnote{a}}&\multirow{2}{*}{\bf{$C_\mathrm{out}$}\tnote{a}}&\multirow{2}{*}{\bf{Method}} &
			\multirow{2}{*}{\bf{MD}\tnote{b}}& \multirow{2}{*}{\bf{SAB}\tnote{b}}&
			\multirow{2}{*}{\bf{DCD}\tnote{b}}&\multicolumn{3}{|c}{\bf{Liver}}&\multicolumn{3}{|c}{\bf{Tumor}}
			\\

			\cline{7-12}
			& & & & & &Dice per case [\%] &Dice global [\%] & VOE &Dice per case [\%]    &Dice global [\%]  &VOE\\
			\hline
			1 &1 &\multicolumn{1}{l|}{$(a)$ Baseline}  & & & &$95.27\pm1.90$  &95.39  &$0.097\pm0.032$ &$57.29\pm33.03$  &67.79  &$0.608\pm0.311$\\           \hline
			5 &1 &\multicolumn{1}{l|}{$(b)$ Baseline}  & & & &$95.34\pm1.78$  &95.56  &$0.087\pm0.033$ &$60.02\pm29.78$  &69.67
			&$0.564\pm0.289$\\           \hline
			7 &1 &\multicolumn{1}{l|}{$(c)$ Baseline}  & & & &$95.39\pm1.76$  &95.59  &$0.084\pm0.032$ &$60.97\pm28.97$	&70.15	&$0.561\pm0.279$\\           \hline

			% 9 &1 &\multicolumn{1}{l|}{$(d)$Baseline}    &- &- &-
			% &95.3 &95.9 &62.5 &68.5\\
			% \hline

			&    &\multicolumn{1}{l|}{$(d)$ Baseline ($1\times$)} & & & &$96.08\pm1.89$	&96.56	&$0.077\pm0.034$ &$62.54\pm28.33$	&72.81	&$0.530\pm0.285$\\
			&    &\multicolumn{1}{l|}{$(e)$ Baseline ($3\times$)} & & & &$96.15\pm1.75$	&96.59	&$0.076\pm0.032$ &$62.67\pm27.80$	&71.25	&$0.531\pm0.276$\\
			& &\multicolumn{1}{l|}{$(f)$ MD}    &\checkmark & & &$\textbf{96.27}\pm\textbf{1.78}$	&\textbf{96.74}	&$0.074\pm0.032$ &$64.62\pm27.29$	&72.48	&$0.510\pm0.269$\\
			5 & 3 &\multicolumn{1}{l|}{$(g)$ MD+SAB}   &\checkmark &\checkmark & &$96.12\pm1.79$	&96.58 &$0.077\pm0.032$ &$66.04\pm23.90$	&75.06	&$0.503\pm0.253$\\
			& &\multicolumn{1}{l|}{$(h)$ MD+DCD}  &\checkmark & &\checkmark &$96.13\pm1.92$	&96.61	&$0.076\pm0.035$ &$65.71\pm23.60$ &71.98	&$0.508\pm0.253$\\
			& &\multicolumn{1}{l|}{$(k)$ MD+SAB+DCD}  &\checkmark &\checkmark &\checkmark &$96.09\pm1.80$	&96.55	&$0.077\pm0.033$ &$67.07\pm23.79$ &73.82	&$0.491\pm0.247$\\
			% &    &\multicolumn{1}{l|}{$(i)$3D-D+DCD} &- &- &- &95.8 &96.0 &62.3 &69.1\\
			% &    &\multicolumn{1}{l|}{$(j)$3D-DN+DCD} &- &- &- &96.2 &96.3 &65.8 &73.2\\
			\hline

			& &\multicolumn{1}{l|}{$(l)$ Baseline ($1\times$)} & & & &$95.69\pm3.08$	&96.51	&$0.084\pm0.034$ &$63.04\pm25.00$	&72.25	&$0.534\pm0.266$\\
			& &\multicolumn{1}{l|}{$(m)$ Baseline ($5\times$)} & & & &$95.97\pm1.82$ &96.44	&$0.079\pm0.033$ &$64.60\pm26.33$ &73.66	&$0.513\pm0.258$\\
			& &\multicolumn{1}{l|}{$(n)$ MD} &\checkmark & & &$96.15\pm1.88$	&96.64	&$0.076\pm0.034$ &$65.69\pm22.65$ &70.67	&$0.511\pm0.245$\\
			7& 5 &\multicolumn{1}{l|}{$(o)$ MD+SAB}  &\checkmark &\checkmark & &$96.20\pm1.67$	&96.60 	&$\textbf{0.074}\pm\textbf{0.030}$ &$66.73\pm24.34$	&74.40 	&$0.494\pm0.241$\\
			&  &\multicolumn{1}{l|}{$(p)$ MD+DCD}  &\checkmark & &\checkmark    &$96.15\pm1.92$	&96.65	&$0.076\pm0.035$ &$67.65\pm21.00$	&72.82	&$0.492\pm0.232$\\
			&  &\multicolumn{1}{l|}{$(q)$ MD+SAB+DCD}  &\checkmark &\checkmark &\checkmark    &$95.95\pm1.96$	&96.50 	&$0.080\pm0.035$ &$\textbf{70.17}\pm\textbf{18.06}$	&\textbf{75.84}	&$\textbf{0.467}\pm\textbf{0.212}$\\
			% &    &\multicolumn{1}{l|}{$(p)$3D-D+DCD} &- &- &- & 96.1 & 96.3 & 66.6 & 73.2\\
			% &    &\multicolumn{1}{l|}{$(q)$3D-DN+DCD} &- &- &- &\textbf{96.4} &\textbf{96.4} &69.0&73.3\\
			\hline
			\toprule[2pt]
		\end{tabular}

		\makeatletter\def\TPT@hsize{}\makeatletter

		\begin{tablenotes}
			\scriptsize
			\item[a] \bf{$C_\mathrm{in}$} and \bf{$C_\mathrm{out}$} represent the number of input slices and output slice predictions, respectively.
			\item[b] \bf{MD}, \bf{SAB} and \bf{DCD} represent the Multi-branch Decoder, the Slice-centric Attention Block and the Densely Connected Dice loss, respectively.
		\end{tablenotes}
	\end{threeparttable}
    \end{adjustbox}

\end{table*}

In the test phase, a sliding-window approach is employed to predict the segmentation mask for an input volume.
Concretely, we extract consecutive multi-slice inputs from the volume by moving along the out-of-plane direction with a stride of one, and predict the segmentation mask for each of these multi-slice inputs.
Therefore, each slice may appear in multiple multi-slice inputs due to the overlap and be predicted multiple times.
The final segmentation mask of a slice is then obtained by averaging its multiple predictions.
Finally, the segmentation masks of all the slices are stacked in sequence to form the segmentation result of the entire volume, which is then resampled to the original resolution, if necessary.

The Dice-per-case score and Dice-global score are adopted in the LiTS challenge as the evaluation metrics to measure the liver and liver tumor segmentation performance. The Dice-per-case score reflects the averaged Dice score across all patients, whereas the Dice-global score is the Dice score evaluated by stacking all volumes into one long volume.
For comparison with other methods, we also use Volume Overlap Error (VOE), Relative Volume Difference (RVD), Average Symmetric Surface Distance (ASSD), Maximum Surface Distance (MSD), and Root Means Square Symmetric Surface Distance (RMSD) as metrics for complementary evaluation.
For interpretation of these evaluation metrics, readers can refer to \cite{bilic2019liver}.
% The Dice score is defined as:
% \begin{equation}
% \label{eq:dice3}
% Dice(G, P) = \frac{2TP}{FP+2TP+FN} = \frac{2|G \cap P|}{|G|+|P|},
% \end{equation}
% where $G$ represents the ground truth; $P$ represents the prediction result; $TP, FP,$ $FN$ denote the number of true positives, false positives, and false negatives, respectively.

\subsubsection{Ablation Study on LiTS}
To analyze the effectiveness of our approach, we conduct ablation studies on a validation dataset consisting of 25 volumes, which are randomly selected from the training set. Since our network backbone is derived from DeepLabv3+~\cite{chen2018encoder}, we take it as the baseline benchmarking method.
We notice that the meta-information about slice thickness and pixel spacing of some cases in the training and validation sets provided by the LiTS organizers is wrong, which makes the values of surface-based metrics in the ablation study weird compared to those presented in Table \ref{tab2}.
Therefore, we do not present surface-based metrics in Table \ref{tab1} for the ablation study.

\paragraph{The number of input slices}
Objects in adjacent slices usually have intrinsic relations in various properties, such as shape and location. % However, 3D isotropic convolution kernels may not be the optimal choice to capture inter-slice information in an anisotropic volume.
In this sense, we employ a 2.5D network, which takes a few adjacent slices as the input to capture the inter-slice information. To verify that the inter-slice information is useful in the segmentation task, we conduct experiments with different numbers of input slices. In Table \ref{tab1}, rows $(a), (b), (c)$ show the results of DeepLabv3+, with $1, 5, 7$ adjacent slices as input, respectively, and the corresponding central slice as output.
We can observe consistent improvement when the number of input slices increases, confirming our assumption that more adjacent slices can provide more inter-slice information for achieving higher segmentation performance.

\paragraph{Effectiveness of the multi-branch decoder}
To verify the effectiveness of the proposed multi-branch decoder, a straightforward baseline is a single-branch decoder that outputs the same number of channels as the multi-branch decoder.
The multi-branch decoder consists of $C_\mathrm{out}$ parallel branches, each having the same structure with the single-branch decoder in the baseline, as shown in Fig. \ref{fig:single-branch-vs-multi-branch}(b). Here, we denote the multi-branch structure as MD.
We present two different settings of the numbers of input slices and output predictions ($C_\mathrm{in}=5, C_\mathrm{out}=3$ and $C_\mathrm{in}=7, C_\mathrm{out}=5$) to verify that our multi-branch design can bring consistent improvement under different inter-slice context. The results are shown in Table \ref{tab1}.
As we can see, compared to the baseline with the same input and output settings (rows $(d)$ and $(l)$), our proposed multi-input multi-output architecture prominently enhances the segmentation of liver and liver tumor, with improvements in both Dice-per-case and Dice-global (rows $(f)$ and $(n)$).
Besides, to further demonstrate that the improvements upon the single-branch decoder are not due to the increased parameters, we also present results of the single-branch decoder with $3\times$ and $5\times$ channels in rows $(e)$ and $(m)$.
We find that even with more parameters, the single-branch decoder still performs worse than our multi-branch decoder in both liver and tumor segmentation.
The multi-branch decoder focuses on modeling the slice-specific information from the mixed volumetric semantic information, which is of great significance for network training with anisotropic data.

\paragraph{Effectiveness of the slice-centric attention block}
Next, we verify the effectiveness of the slice-centric attention block, which is abbreviated as SAB in Table \ref{tab1}.
Rows $(g)$ and $(o)$ show the detailed results with the multi-branch decoder and SAB for liver and liver tumor segmentation.
Compared to $(f)$ and $(n)$, which are equipped only with the multi-branch decoder, the SAB achieves $1.42\%$ and $1.04\%$ improvements in Dice-per-case for tumor segmentation.
The improvements demonstrate that the deep semantics learned by the encoder contains vast amounts of redundant information, and directly decoding with convolution operators cannot effectively extract slice-specific features.
With the proposed attention mechanism, the network can steer the allocation of available semantic features towards the most informative components for different slices, thus bringing improvements in terms of accuracy.

\begin{figure}[!t]
	\begin{center}
		\includegraphics[width=0.9 \columnwidth]{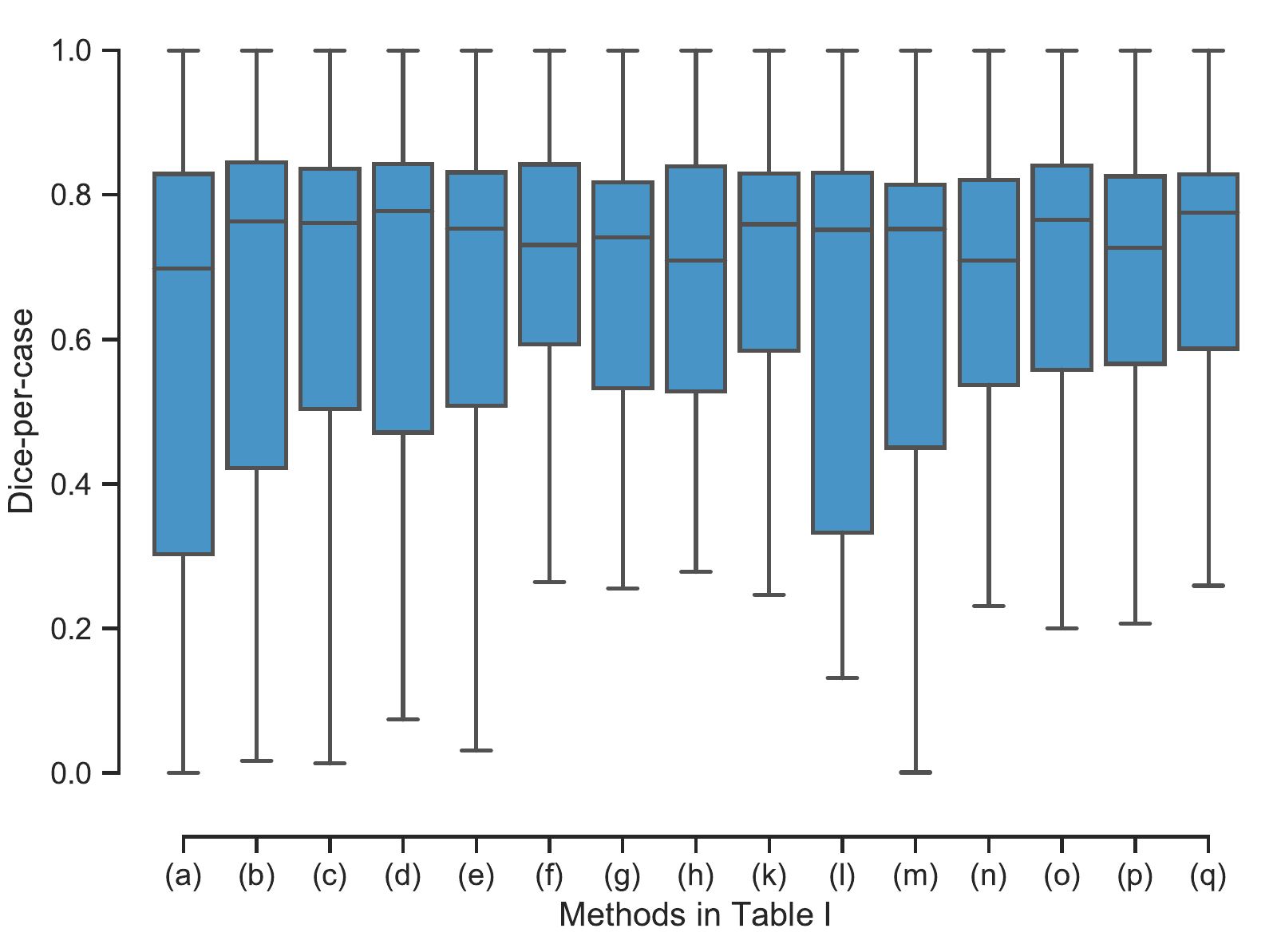}
	\end{center}
	\caption{Box-plots of each method in Table \ref{tab1} on tumor segmentation. Horizontal axis represents the methods listed in Table \ref{tab1}, and vertical axis represents the Dice-per-case score.}
	\label{boxplot}
\end{figure}

\begin{table*}[!t]
	\caption {The segmentation results of baseline $(d)$ and baseline $(l)$ in TABLE \ref{tab1} with unified resolution on the LiTS validation dataset.} \label{tab_sambd_ur}
	\centering
		\begin{threeparttable}
			\begin{tabular}{c c|c|c c c|c c c}
				\toprule[2pt]
				\multirow{2}{*}{\bf{$C_\mathrm{in}$}\tnote{a}}&\multirow{2}{*}{\bf{$C_\mathrm{out}$}\tnote{a}}&\bf{Unified Resolution} &\multicolumn{3}{c|}{\bf{Liver}}&\multicolumn{3}{c}{\bf{Tumor}}
				\\

				\cline{4-9}
				& & [mm] &Dice per case [\%] &Dice global [\%] &VOE &Dice per case [\%]    &Dice global [\%]  &VOE\\
				\hline
				 & &\multicolumn{1}{c|}{2}  &$91.02\pm7.21$	&92.43	&$0.158\pm0.103$ &$46.42\pm32.67$   &64.18  &$0.636\pm0.297$\\
				5 &3 &\multicolumn{1}{c|}{1}  &$95.41\pm1.80$	&95.74	&$0.088\pm0.032$ &$63.80\pm24.27$ &72.14	&$0.440\pm0.260$\\
				 & &\multicolumn{1}{c|}{0.75} &$95.25\pm2.07$	&95.79	&$0.090\pm0.037$&$63.42\pm24.89$	&68.02	&$0.489\pm0.263$\\
				\hline
				 & &\multicolumn{1}{c|}{2}  &$95.11\pm1.85$	&95.45	&$0.093\pm0.033$ &$55.89\pm32.17$   &72.05	&$0.503\pm0.310$\\
				7 &5 &\multicolumn{1}{c|}{1}  &$95.66\pm1.95$	&96.17	&$0.083\pm0.035$ &$63.75\pm25.36$	&72.53	&$0.486\pm0.255$\\
				 & &\multicolumn{1}{c|}{0.75}  &$95.27\pm1.95$	&95.78	&$0.090\pm0.035$ &$65.52\pm24.06$	&69.87	&$0.468\pm0.254$\\

				\hline
				\toprule[2pt]
			\end{tabular}

			\makeatletter\def\TPT@hsize{}\makeatletter

			\begin{tablenotes}
				\scriptsize
				\item[a] \bf{$C_\mathrm{in}$} and \bf{$C_\mathrm{out}$} represent the number of input slices and output slice predictions, respectively.
			\end{tablenotes}
		\end{threeparttable}

\end{table*}

\begin{table*}[!t]
	\caption {The segmentation results of 3D U-Net \cite{cciccek20163d} with unified resolution on the LiTS validation dataset.} \label{tab_unet_ur}
	\centering
	% \begin{adjustbox}{width=0.8\textwidth}

			\begin{tabular}{c|c c c| cc c c}
				\toprule[2pt]
				\bf{Unified Resolution} &\multicolumn{3}{c}{\bf{Liver}}&\multicolumn{3}{c}{\bf{Tumor}}	\\

				\cline{2-7}
				[mm] &Dice per case [\%] &Dice global [\%] &VOE &Dice per case [\%]    &Dice global [\%]  &VOE\\
				\hline
				\multicolumn{1}{c|}{2}  &$90.95\pm7.05$	&92.34	&$0.160\pm0.102$ &$40.43\pm35.03$	&63.34	&$0.433\pm0.364$\\
				\multicolumn{1}{c|}{1}  &$93.55\pm3.19$	&94.27	&$0.120\pm0.054$ &$59.06\pm27.79$	&65.29	&$0.529\pm0.272$\\
				\multicolumn{1}{c|}{0.75} &$95.41\pm1.79$	&95.77	&$0.087\pm0.319$ &$62.49\pm27.27$ &71.80	&  $0.492\pm0.275$\\
				\hline
				\toprule[2pt]
			\end{tabular}

	% \end{adjustbox}

\end{table*}

\begin{table}[!t]
	\color{red}
	\caption {P-values for the paired t-tests between our method and the single-branch decoder that outputs the same number of channels as the multi-branch decoder.} \label{tab_pvalues}
	\centering

	\begin{adjustbox}{width=\linewidth}
	\begin{tabular}{cc|cc}
   \toprule

	$C_\mathrm{in}=5, C_\mathrm{out}=3$ & Dice-per-case & $C_\mathrm{in}=7, C_\mathrm{out}=5$ &  Dice-per-case \\
	\midrule
	p-value & 0.034 & p-value & 0.010  \\
	\bottomrule
	\end{tabular}
	\end{adjustbox}

\end{table}

\paragraph{Effectiveness of the DCD loss}
Besides the regular Dice loss which measures the difference between the segmentation prediction and given ground truth mask, the method is further equipped with a DCD loss as shown in Eq. (\ref{eq:dice4}), which is proposed to regularize the inter-slice predictions to be more coherent in the label space. To investigate the effectiveness of the DCD loss, we conduct several ablation studies. First, we train the network with the multi-branch decoder using only the DCD loss, as shown in Table \ref{tab1}, rows $(h)$ and $(p)$. Compared to training with the Dice loss (rows $(f)$ and $(n)$ in Table \ref{tab1}), the DCD loss improves the overall Dice-per-case score for tumor by
$1.09\%$ and $1.96\%$, respectively. Then, we add the slice-centric attention block to the network (rows $(k)$ and $(q)$ in Table \ref{tab1}) and train it with the Dice loss and the DCD loss, respectively. We get a $1.03\%$ improvement in Dice-per-case score for tumor with 5-slice input and 3-slice output, while $3.44\%$ improvement with 7-slice input and 5-slice output. It is worth mentioning that compared to the baseline, the proposed method performs preferably in segmenting tumor with an improvement of about $7.13\%$ in Dice-per-case score by comparing rows $(l)$ and $(q)$, which proves the effectiveness of the proposed network with all modules enabled.

\paragraph{Statistical analysis}
To analyze whether the performance improvement of the proposed SAMBD is statistically significant, we conduct the paired t-test as \cite{liu2020ms} on tumor segmentation between our method and the single-branch decoder that outputs the same number of channels as the multi-branch decoder \textcolor{blue}{with two configurations of $C_\mathrm{out}$ and $C_\mathrm{in}$.
We evaluate the significance for Dice-per-case with a significance level of $0.05$.
The p-values are shown in Table \ref{tab_pvalues}.
As we can see, all the p-values are below 0.05, demonstrating that our improvements upon the single-branch decoder are statistically significant.}

\paragraph{Box-plots}
In Fig. \ref{boxplot}, we show the box-plots of each method listed in Table \ref{tab1} in terms of Dice-per-case on tumor segmentation.
Compared to other methods (rows $(a)-(h)$ and $(l)-(p)$), the proposed SAMBD presents results that not only have higher median accuracy but also show less dispersion, indicating consistently better performance in general.

% \paragraph{3D variant}
% We replace the 2.5D multi-branch decoder with its 3D corresponding variant introduced in Section \ref{D} to prove the necessity of learning slice-aware information for anisotropic data. The 3D-D in Table \ref{tab1} presents decoder with isotropic 3D convolution, namely $3 \times 3 \times 3$ kernels, while the 3D-DN presents decoder with anisotropic 3D convolution and non-local block. The 3D-DN architecture achieves a Dice-per-case of $65.8\%$ for tumor with 5-slice input and 3-slice output (row $(j)$), while $69.0\%$ with 7-slice input and 5-slice output (row $(q)$), achieving $3.5\%$ and $2.4\%$ improvement compared to 3D-D architecture shown in rows $(i)$ and $(p)$, respectively. The results confirm that our motivation to learn slice-aware information from volumetric features is necessary.

\begin{table*}[!t] \caption {Comparison with state-of-the-art methods for liver and tumor segmentation on the LiTS test dataset.} \label{tab2}
	\centering
	\begin{adjustbox}{width=\textwidth}
	\begin{tabular}{c|c|c|c|c|c|c|c|c|c|c|c|c|c|c}
		\toprule[2pt]
		\multirow{2}{*}{\bf{Method}} & \multicolumn{7}{|c}{\bf{Liver}}&\multicolumn{7}{|c}{\bf{Tumor}}\\
		\cline{2-15}
		& Dice per case [\%] & Dice global [\%] & VOE & RVD & ASSD & MSD & RMSD & Dice per case [\%] &Dice global [\%] & VOE & RVD & ASSD & MSD & RMSD\\
		\hline

		Han \cite{han2017automatic} & 96.0 & 96.5 & 0.077 & -0.004 & 1.15 & 24.499 & 2.421 & 67.6 & 79.6 & 0.383 & 0.464 & 1.143 & 7.322 & 1.728\\
		3D DenseUNet \cite{li2018h} & 93.6 & 92.9 & - & - & - & - & - & 59.4 & 78.8 & - & - & - & - & -\\
		H-DenseUNet \cite{li2018h}  & 96.1 & 96.5 & 0.074 & -0.018 & 1.45 & 27.118 & 3.15 & 72.2 & 82.4 & 0.366 & 4.272 & \textbf{1.102} & 6.228 & \textbf{1.595}\\
		AH-Net \cite{liu20183d}     & 96.3 & 97.0 & 0.07 & -0.004 & 1.099 & 23.992 & 2.398 & 63.4 & \textbf{83.4} &\textbf{0.353} & 0.365 & 1.185 & 6.482 & 1.667\\
		DeepX \cite{yuan2017hierarchical}  & 96.3 & 96.7 & 0.071 & -0.01 & 1.104 & 23.847 & 2.303 & 65.7 & 82.0 & 0.378 & 0.288 & 1.151 & 6.269 & 1.678\\
		\hline
		SAMBD                       & 96.5 & 97.0 & \textbf{0.065} & 0.004 & 0.971 & 21.997 & 2.034 & 72.8 &81.0 &0.405 & -0.208 & 1.258 & 6.582 & 1.796\\
		SAMBD (ensemble)        &\textbf{96.6} & \textbf{97.1} & \textbf{0.065} & \textbf{0.002} & \textbf{0.953} & \textbf{21.933} & \textbf{1.998} &\textbf{73.6} & 81.2 & 0.401 & \textbf{-0.196} & 1.174 & \textbf{6.18} & 1.675\\
		\hline
		\toprule[2pt]
	\end{tabular}
	\end{adjustbox}
\end{table*}

\begin{figure*}[t!]
\begin{center}
\includegraphics[width=0.85 \linewidth]{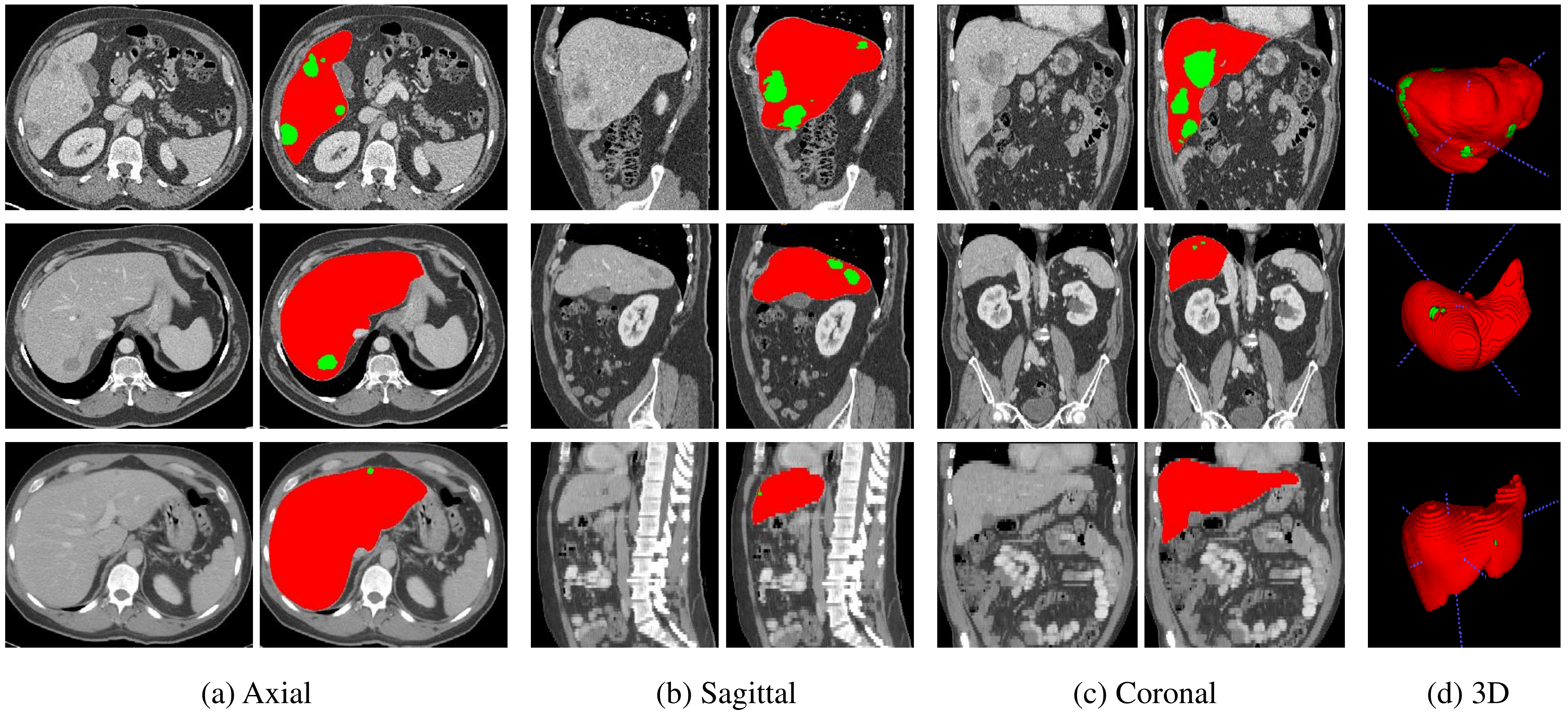}
\end{center}
\caption{Example segmentation results on the LiTS test set. Each row shows a CT scan acquired from an individual subject with different slice thickness (0.7 $mm$, 2.5 $mm$, and 5 $mm$ from top to bottom, respectively). The first six columns show the original CT scans and corresponding segmentation results in the axial, sagittal, and coronal planes, respectively. The last column shows segmentation results in a 3D view. Red represents the liver and green the liver tumor.}
\label{test_result_lits}
\end{figure*}

\subsubsection{Does a unified resolution help address the data variability problem?}
In the medical image segmentation, a widely adopted approach to the data variability problem is resampling the data into a unified resolution.
We present the results of the single-branch decoder that outputs the same number of channels as the multi-branch decoder with different unified resolutions in Table \ref{tab_sambd_ur}.
Compared to the results of SAMBD in rows $(k)$ and $(q)$ of Table \ref{tab1}, the results with the single-branch decoder still present a large performance gap in Dice-per-case, Dice-global, and VOE.

Since 3D networks are good at extracting contextual information and the resampling operation is widely adopted to process data suffering large variations, we present extra experimental results with 3D U-Net \cite{cciccek20163d} to see if it can produce sound results with a unified resolution.
The results are shown in Table \ref{tab_unet_ur}.
From the table, we can observe that 3D U-Net produces significantly different results with different input resolutions.
The best performance is achieved at the highest resolution ($0.75\times0.75\times0.75$ in mm), yet substantially lower than the results of SAMBD in row $(q)$ of Table \ref{tab1} (\textit{e.g.,} 62.49\% vs. 70.17\% in Dice-per-case for tumor segmentation).

The superior performance of the proposed SAMBD towards 2D/3D networks with a unified resolution verifies the effectiveness of our slice-aware design in processing anisotropic data.

\subsubsection{Comparison with Other Methods}
Based on the results from the ablation study, we pick the best network architecture and hyper-parameters, and train the model on the whole LiTS training set with 131 CT scans. We evaluate the model on 70 test cases and submit the results to the challenge website. Table \ref{tab2} tabulates the quantitative comparison results of our proposed SAMBD and several state-of-the-art methods already published. All of these top-ranking methods employ deep learning based approaches, demonstrating the effectiveness of CNNs in the field of medical image analysis. Han \cite{han2017automatic} adopted a 32-layer U-Net-alike architecture, where adjacent slices were employed as input and produced segmentation maps through a single-branch decoder. Li \textit{et al.} \cite{li2018h} and Liu \textit{et al.} \cite{liu20183d} transferred convolutional features learned from 2D images to 3D volumes and then applied 3D convolutional kernels to extract 3D context. Yuan \cite{yuan2017hierarchical} developed a hierarchical framework to segment the liver and tumor in three steps. Compared to these methods, we are the first who pay attention to the large inconsistency between the in-plane pixel spacing and out-of-plane slice thickness.

As shown in Table \ref{tab2}, our method achieves the best segmentation accuracy for both liver and tumor in the Dice-per-case score even with a single model.
Using an ensemble version, we achieve state-of-the-art performances in three of the four main evaluation metrics, including the Dice-per-case score of the tumor, and Dice-global and Dice-per-case scores of the liver, demonstrating the effectiveness of our method.
Besides, our methods show very competitive results on the complementary evaluation metrics, achieving the best results in VOE, RVD, ASSD, MSD, and RMSD for liver, as well as best results in RVD and MSD for tumor.
Fig. \ref{test_result_lits} shows several examples of the segmentation results with different slice thickness (0.7 $mm$, 2.5 $mm$, and 5 $mm$, respectively).
By taking into consideration the information asymmetry along with the in- and out-of-plane directions into our network design, our method presents decent ability in segmenting liver and liver tumors across a wide range of resolution settings.

\begin{figure}[!t]
\begin{center}
	\includegraphics[width=1.\columnwidth]{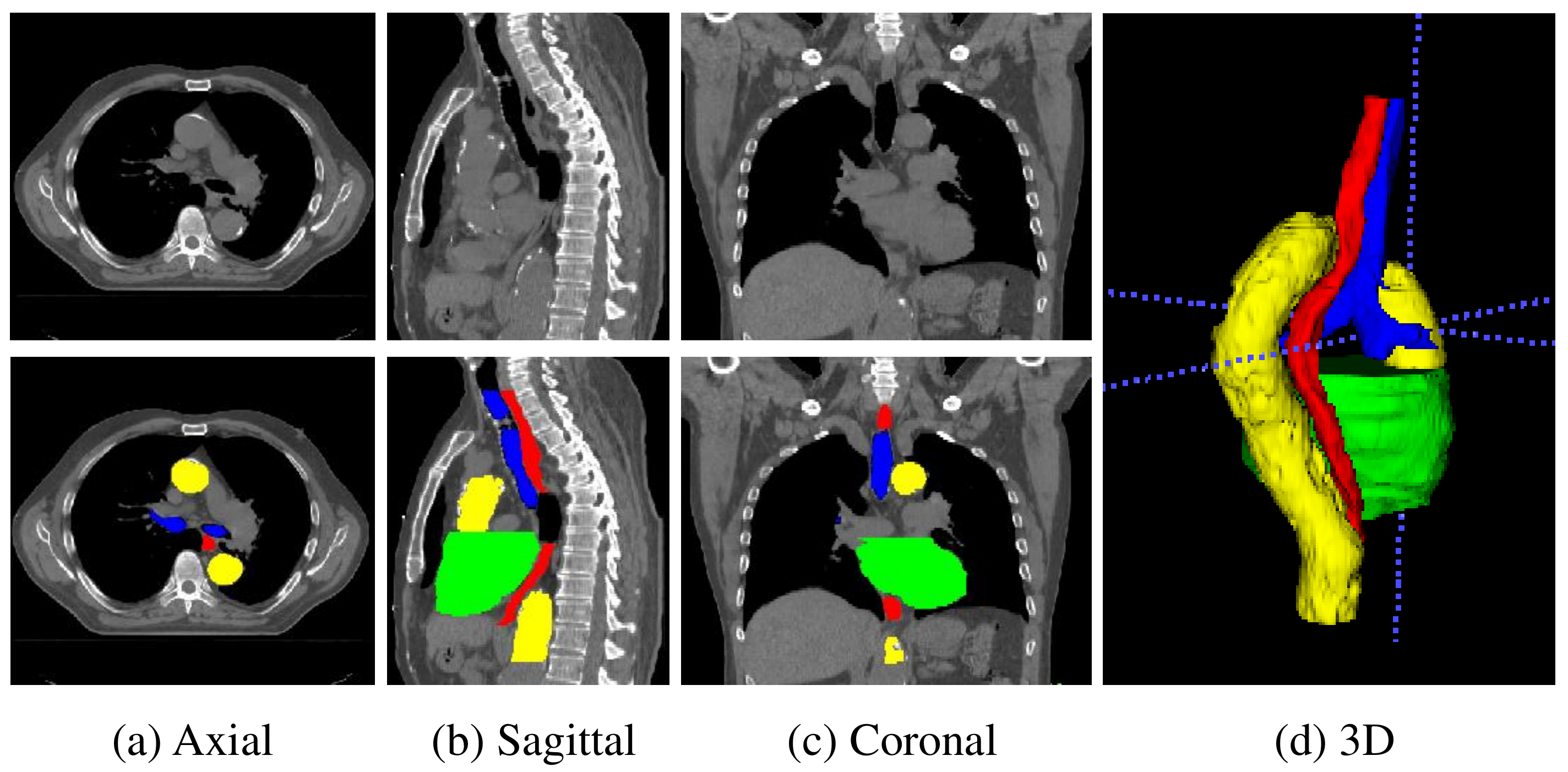}
\end{center}
\caption{A typical CT scan in the SegTHOR dataset and the corresponding segmentation ground truth with red for the esophagus, green for the heart, blue for the trachea, and yellow for the aorta.}
\label{segthor}
\end{figure}

\subsection{Experiments on SegTHOR} \label{exp_segthor}
\subsubsection{SegTHOR}

\begin{table*}[!t] \caption {Segmentation results on the test dataset in leaderboard of 2019 SegTHOR Challenge (until May 10, 2019 when the live challenge of ISBI ended.)} \label{tab3}
	\centering
	\begin{tabular}{c|c|c|c|c|c|c|c|c|c}
		\toprule[2pt]
		\multirow{2}{*}{\bf{Rank}}&\multirow{2}{*}{\bf{User/Method}} & \multicolumn{2}{|c}{\bf{Esophagus}}&\multicolumn{2}{|c}{\bf{Heart}}&\multicolumn{2}{|c}{\bf{Trachea}}&\multicolumn{2}{|c}{\bf{Aorta}}
		\\
		\cline{3-10}
		& & Dice [\%]    &Hausdorff  &Dice [\%]   &Hausdorff &Dice [\%]   &Hausdorff &Dice [\%]   &Hausdorff\\
		\hline

		1&gaoking132         & 86.51   &0.2590  &95.36  &0.1272		&\textbf{92.76}		&\textbf{0.1453}		&94.64		&0.1209 \\
		2&MILab~\cite{hemulti}     &85.94    &0.2743  &95.00  &0.1383		&92.01		&0.1824		&\textbf{94.84}		&\textbf{0.1129}\\
		3&Jone      &85.91    &0.3185  &94.89  &0.1435  &92.19		&0.1590 	&94.73	 &0.1251\\
		4&hyang         &83.81    &0.3534  &\textbf{95.42}  &\textbf{0.1208}		&92.33		&0.1973		&94.43		&0.1290\\
		5&dp  &83.39    &0.3351  &95.19  &0.1325	&91.57	&0.2041	&93.51	&0.1980\\
		6&grr  &85.82    &0.2928  &94.56  &0.1867	&91.53	&0.2090	&93.91	&0.2010\\
		7&dlachinov  &85.72    &0.2686  &94.36  &0.1761	&91.55	&0.1736	&91.82	&0.2214\\
		8&ZWB  &82.03    &0.3838  &94.58  &0.1594	&92.17	&0.2045	&94.33	&0.1551	\\
		9&svesal~\cite{vesal20192d}  &85.79   &0.3303	&94.15 & 0.2263 &92.57  &0.1929  &93.75	&0.2971	\\
		10&Louisvh  &83.61   &0.3399	&94.02 &0.1973 &90.68  &0.2091  &93.32	&0.2732	\\
		\hline
		-& Chen \textit{et al.} \cite{chentwo}         & 81.66   &0.4914  &93.29  &0.2417		&89.10		&0.2746	&92.32		&0.3081 \\
		-& Zhang \textit{et al.} \cite{zhangsegmentation} &77.32 &1.6774 &93.84 &0.2089 &89.39 &0.2741 &92.32 &0.3081 \\
		\hline
		2&SAMBD  &\textbf{86.57}    &\textbf{0.2478}  &95.28  &0.1296	&92.56	&0.1817	&94.43	&0.1498\\
		\hline
		\toprule[2pt]
	\end{tabular}
\end{table*}

\textcolor{blue}{The dataset}\footnote{\color{blue}https://competitions.codalab.org/competitions/21145} \textcolor{blue}{\cite{trullo2019multiorgan}} is provided by ISBI 2019, with the purpose of addressing the problem of Organs at Risk (OAR) segmentation in CT images. SegTHOR focuses on four OARs: heart, aorta, trachea, and esophagus (refer to Fig. \ref{segthor} for data visualization), and pays particular attention to esophagus segmentation (the esophagus is the most challenging to segment due to its variable location relative to neighboring organs and low-intensity contrast to background). All CT images are obtained from 60 patients (11,084 slices) with non-small cell lung cancer and have already been randomly split into a training set with 40 volumes (7,390 slices) and a test set with 20 volumes (3,694 slices). The planar size is $512 \times 512$ in pixels and the in-plane pixel spacing varies from $0.9$ $mm$ to $1.37$ $mm$, while the out-of-plane slice thickness varies from $2$ $mm$ to $3.7$ $mm$. All the organs are delineated by an experienced radiation oncologist.

\subsubsection{Implementation Details and Evaluation Metrics}
We set the HU value range to $[-1000, 500]$ to exclude irrelevant organs and objects. The slice thicknesses of all subjects are resampled to $2.5$ $mm$.
Hyperparameters and data augmentation are the same as those used in the experiments on the LiTS dataset.
In addition, there are great variations of shape and position among four organs: some extend along the out-of-the plane direction (\textit{e.g.}, esophagus) and some have a large volume (\textit{e.g.}, heart). To address these issues, we cut the 3D CT scans into slices along the axial, sagittal, and coronal plane, respectively, and then feed them into the network together.
The size of the axial slices is $512 \times 512$ pixels, whereas those of the sagittal and coronal slices are $512 \times z$ and $z \times 512$ pixels ($z$ is the number of axial slices), respectively.
To facilitate the segmentation network, the sagittal and coronal slices are then resized to $512 \times 512$ pixels.
In the test phase, we separately predict the segmentation masks in three orientations (i.e., the axial, sagittal, and coronal) for each volume in a way similar to the experiments on the LiTS dataset,  and ensemble them with the majority voting to produce the final segmentation mask.

The overlap Dice metric and the Hausdorff distance \cite{vesal20192d} are used by the SegTHOR challenge as the official evaluation metrics.
The Dice score and Hausdorff distance are complementary metrics and our aim is to make the Hausdorff distance close to 0 while the Dice score close to 1.

% Among them, Hausdorff measures the maximum degree of mismatch between A and B, which is defined as:
% \begin{equation}
% \label{eq:Hausdorff}
% H(A, B) = max(h(A, B), h(B, A))
% \end{equation}
% where $h(A, B) = \max \limits_{a \in A} \min \limits_{b \in B} \left \| a-b \right \|$, and $h(B, A) = \max \limits_{b \in B} \min \limits_{a \in A} \left \| b-a \right \|$. $h(A, B)$ and $h(B, A)$ are the so-called directed Hausdorff distance.

\subsubsection{Comparison with Other Methods}
In this experiment, we pay special attention to the segmentation of the esophagus, because of its hard-to-distinguish boundary and low contrast. Table \ref{tab3} lists the performance of the top 10 teams on the leaderboard. Our method outperforms other methods on the segmentation of the esophagus and achieves very competitive performance for heart and trachea segmentation. We finally achieve the tied second place by the overall rank and the first place on esophagus segmentation.
It is worth mentioning that our method is not specially tailored for the tasks of SegTHOR---unlike the contrasting methods in Table \ref{tab3} which were intended for the specific challenge---yet produces such competitive results, demonstrating its generalizability.

% \begin{figure*}[!t]
% \begin{center}
% 	\includegraphics[width=0.8\linewidth]{pics/segthor_result.pdf}
% \end{center}
% \caption{Examples of segmentation results on the SegTHOR test set. Each row shows a CT scan acquired from an individual subject. The first six columns show the original CT scans and corresponding segmentation results in the axial, sagittal, and coronal plane, respectively. The last volume shows segmentation results in a 3D view.}
% \label{segthor_result}
% \end{figure*}

Among all the methods listed in Table \ref{tab3}, He \textit{et al.} \cite{hemulti} developed a uniform U-like encoder-decoder architecture for the segmentation of thoracic organs, which combined the major task of local pixel-wise segmentation and an auxiliary task of global slice classification. Vesal \textit{et al.} \cite{vesal20192d} employed a 2D U-Net combined with dilated convolutions using only one slice as input. Chen \textit{et al.} \cite{chentwo} and Zhang \textit{et al.} \cite{zhangsegmentation} segmented thoracic organs through a two-stage strategy, where four organs were first localized and then the precise segmentation stage was applied based on the location.
Our method concentrates on making use of the relationships between adjacent slices and learning slice-specific information, which helps identify inconspicuous objects. The segmentation result of the esophagus in Table \ref{tab3} confirms this.
% We show some examples of the segmentation results in Fig. \ref{segthor_result}.
We conclude that our method can conquer the challenge and well segment the OARs from CT scans.

\section{Discussion}\label{discussion}
Accurate segmentation of liver and liver tumors in CT images facilitates the quantitative assessment of the tumor burden, treatment planning, and prognosis. There have been considerable debates over 2D versus 3D networks on 3D medical images---choosing 2D networks for the benefit of 2D pretraining and large-scale training sets or alternatively 3D networks for native 3D representation learning \cite{yang2019reinventing,yu2019thickened}.
This paper innovatively identifies the wide variation in the ratio between intra- and inter-slice resolutions as a crucial obstacle to the performance, which may in turn affect decision-making in choosing 2D or 3D networks.
This argument is also supported by the five-fold cross-validation results of nnU-Net \cite{isensee2018nnu}, in which 2D U-Net produces inferior results on the brain tumour, heart, hippocampus, lung, and pancreas tasks than 3D U-Net, but better results on prostate and liver tasks.
In this sense, we hope this work would provide a different perspective of deciding 2D or 3D networks.

It is common to take multiple slices as input and output a single slice, either for making up a three-channel image for pre-trained weights or pursuing a 3D context \cite{han2017automatic}.
Our work takes a step further to output multiple slices, which enables to explore extra design space for the decoder and loss function for supervision to maintain 3D coherence between slices.
Our focus is then on how to learn the most discriminative features for each individual slice, which avoids directly processing anisotropic information.
To this end, we propose a multi-branch decoder to explicitly re-establish discriminative features for each separate slice by associating each slice with a separate branch.
As far as we know, only few existing works emphasize learning slice-specific features in a 2.5D network.
As verified in the experiment, such a slice-aware design greatly boosts the performance compared to the multi-output single-branch decoder that does not distinguish different slices in the feature learning.
Although in clinical practice, some data may not show large variations, we believe that the slice-aware design would still bring extra help to the segmentation performance.

The limitation of such an explicit design of the multi-branch decoder is that it is computationally prohibitive when the number of output slices becomes large.
In this work, we leverage and embed an attention mechanism into the multi-branch decoder, which is expected to steer the allocation of slice-specific semantic features towards the most informative components for each output slice by fully-exploiting the inter-slice correlations.
We believe that it is novel to adopt such an attention mechanism to strengthen the discriminative power of each slice, yet we acknowledge that there should also be other design choices of the attention block.
We plan to integrate our multi-branch decoder with other well-designed attention blocks (\textit{e.g.}, considering multi-scale features) in future work, seeking more accurate segmentation of the images with slice-aware modeling.
\textcolor{blue}{Besides, in this work we only study the proposed SAMBD with $C_\mathrm{in}=5,\ C_\mathrm{out}=3$ and $C_\mathrm{in}=7,\ C_\mathrm{out}=5$.
Future work should investigate an optimal way to determine the configuration that is general enough for satisfactory results in most cases.
Moreover, although our framework demonstrates competitive computational complexity with the DeepLabV3+ \cite{chen2018encoder} and H-DenseUNet \cite{li2018h} (Table~\ref{tab_complexity}), methodologies for more compact deep neural network designs (e.g., \cite{ghosh2019segfast}) can be considered in future work.}
%Moreover, although our work still achieves competitive computation complexity as demonstrated in Section \ref{F}, a more compact deep neural network can be taken into consideration just as the efforts in \cite{ghosh2019segfast} in future work.

% Although the computation complexity is usually a secondary evaluation criterion in most clinical scenarios

\begin{figure}[!t]
	\begin{center}
		\includegraphics[width=1. \columnwidth]{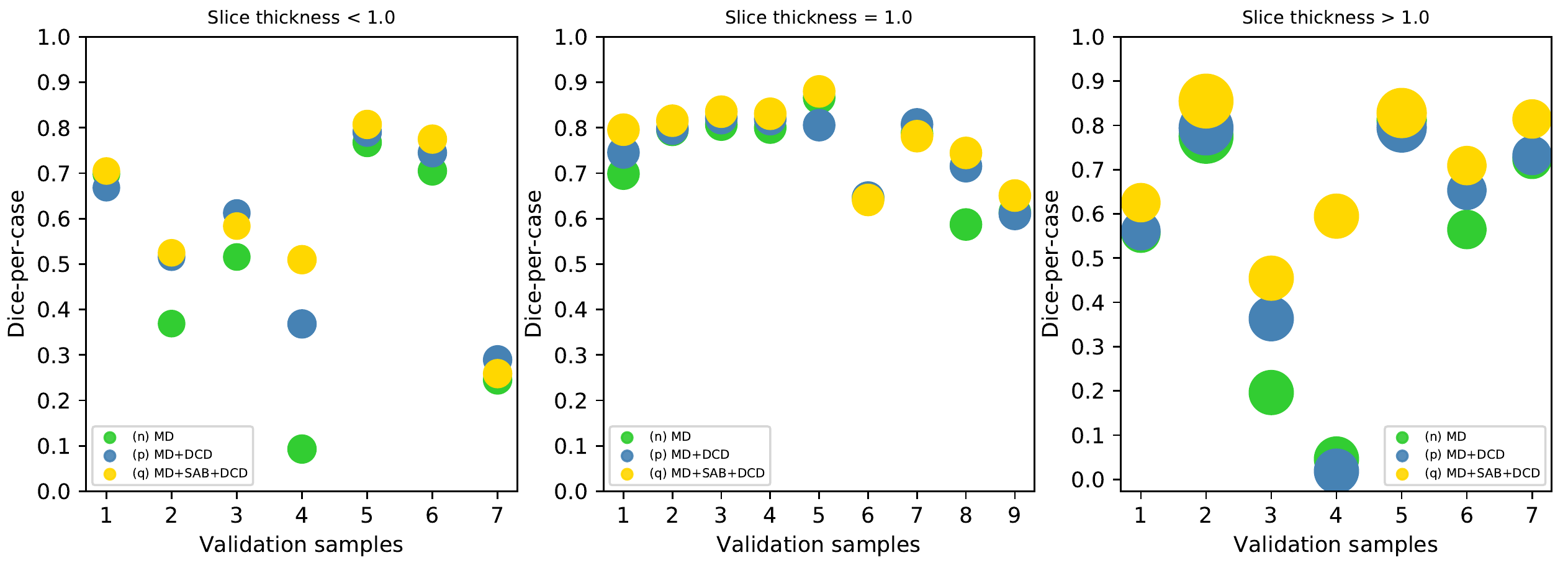}
	\end{center}
	\caption{The segmentation accuracy of different methods in rows (n), (p), and (q) of Table \ref{tab1} grouped by slice thickness less than, eqaul to, greater than 1 $mm$. We find consistent performance improvement brought by the DCD loss irrespective of different slice thicknesses.}
	\label{spacing_acc}
\end{figure}

A potential concern of the DCD loss is that it may potentially affect the proper training when the slice thickness is large.
Actually, the motivation of this regularizer is to supplement inter-slice information and thus improve inter-slice coherence.
There should be a balance between the main loss function and the regularizer just as traditional machine learning algorithms (such as Lasso).
In Fig. \ref{spacing_acc}, we group the segmentation accuracy of different methods (rows $(n)$, $(p)$, and $(q)$ in Table \ref{tab1}) by slice thickness into three groups: less than, equal to, and greater than 1 $mm$.
From the figure, we find consistent performance improvement brought by that DCD loss with different slice thicknesses.
Besides, the range (from 0.69 to 5 in mm) of slice thickness in LiTS is large enough to cover most cases in clinical practice.
Future works should consider the effectiveness of such a regularizer in other segmentation tasks.

Considering that the problem of data variations in resolution is prevalent in medical imaging and we believe that the proposed SAMBD can generalize well to other segmentation tasks, we also evaluated SAMBD on SegTHOR to verify this in Section \ref{exp_segthor}.
% The superior performance---the tied second place by the overall rank and the first place on esophagus segmentation (especially challenging due to its variable locations and low contrast)---proved the robustness and generalizability of the proposed method.
It is noteworthy that, here, we re-trained all over again on the SegTHOR dataset; it should be a potential direction to explore the transfer learning techniques \cite{chen2019med3d} (for example, fine-tuning with pre-trained weights) to speed up the training process.
\textcolor{blue}{Besides, to further verify the effectiveness of our proposed method in conquering data variations in resolution, we plan to test our method on more organs/tumors in other body parts and imaging modalities.}

\section{Conclusion}\label{conclusion}
In this paper, we rethought the debates over 2D versus 3D networks from a data viewpoint, where we identified the wide variation in the ratio between intra- and inter-slice resolutions as an important obstacle to the performance. To circumvent this, we proposed a slice-aware multi-input multi-output structure to emphasize the importance of feature learning for each slice. A multi-branch decoder with a slice-centric attention block was proposed to gradually and explicitly re-establish discriminative features for each slice by fully-exploiting intra- and inter-slice information learned by the encoder with the widely adopted attention mechanism.
To further enhance the correlation between slices and enable coherent segmentation, we proposed a densely connected Dice loss as a regularization term. Quantitative evaluations on the LiTS and SegTHOR datasets demonstrated that our approach could significantly improve segmentation accuracy for anisotropic data.

% or


%\appendices

%\section{}

% \section*{Acknowledgment}

\ifCLASSOPTIONcaptionsoff
  \newpage
\fi



\bibliographystyle{IEEEtran}
\bibliography{reference}
\end{document}